\documentclass[a4paper,3p,number,preprint,review]{elsarticle}

\usepackage{amsmath}
\usepackage{bm}
\usepackage[dvipsnames,svgnames,x11names]{xcolor}

 \usepackage[final]{changes}

\definechangesauthor[color=BrickRed]{rev2}

\newcommand{\ahalf}{\frac{1}{2}}

\newcommand{\dsty}[1]{\displaystyle{#1}}
\newcommand{\sph}[1]{\left\langle {#1} \right\rangle}

\begin{document}

\begin{frontmatter}
  
\title{A geometric formulation of the Shepard renormalization factor }

\author[cehinav,faian]{J.~Calderon-Sanchez\corref{cor1}}
\ead{javier.calderon@upm.es}
  
\author[nasal]{J.L.~Cercos-Pita}
\ead{jl.cercos@upm.es}

\author[cehinav,faian]{D.~Duque}
\ead{daniel.duque@upm.es}


\address[cehinav]{%
  CEHINAV, ETS Ingenieros Navales,
  Universidad Polit\'ecnica de Madrid,
  Avd. de la Memoria 4,
  28040 Madrid, Spain}
\address[nasal]{%
  Research department, NASAL Systems SL, General Orgaz 23, 28020 Madrid, Spain}
\address[faian]{Dep. FAIAN,
  ETSIAE,
  Pza. de Cardenal Cisneros, 3 
  Universidad Polit\'ecnica de Madrid,
  28040 Madrid, Spain}

\begin{abstract}
\label{Abstract}

The correct treatment of boundary conditions is a key step in the development of the SPH method. The SPH community has to face several challenges in this regard --- in particular, a primordial aspect for any boundary formulation is to ensure the consistency of the operators in presence of boundaries and free surfaces.
A new implementation is proposed, based on the existing numerical boundary integrals formulation. A new kernel expression is developed to compute the Shepard renormalization factor at the boundary purely as a function of the geometry. In order to evaluate this factor, the resulting 
expression is split into numerical and analytical parts, which allows accurately computing the Shepard factor.
The new expression is satisfactorily tested for different planar geometries, showing that problems featuring free surfaces and boundaries are solved.
The methodology is also extended to 3-D geometries without great increase in computational cost.
\end{abstract}

\begin{keyword}
Particle methods \sep Meshless methods \sep Smoothed Particle Hydrodynamics \sep Boundary Integrals
\end{keyword}

\end{frontmatter}

\section{Introduction.}
\label{1. Introduction}

The treatment of boundary conditions is an important step in the development of any numerical method. During the past few years, SPH methodology regarding boundary conditions has improved considerably, thanks in part to the community addressing the SPHERIC Grand Challenges, proposed by the SPHERIC SPH Numerical Development Working Group. Boundary conditions issues are specially relevant when dealing with free-surface flows, as discussed in detail in Ref. \cite{Colagrossi09}.

The truncation of the kernel at the ends of the domain makes SPH interpolation process inaccurate, unless consistency of the operators is recovered. In this regard, several approaches to model boundary conditions exist, although two main groups can be distinguished among all the options available: fluid extensions and contour closure.

Fluid extensions are based on the expansion of the system beyond the boundaries, in order to complete the kernel support with mirror images of the fluid particles, which are added to the SPH governing equations in order to restore the consistency of the differential operators. Along this line, the ghost particles approach \cite{Bouscasse13b,Marrone13} has become the most popular option. Nevertheless, some non-trivial challenges should be addressed, such as the choices for the particle locations, which can turn to a hard task for relatively complex geometries, or the procedure to assign field values to the new particles --- see e.g. Refs. \cite{Colagrossi11,Merino13}.

Alternatively, wall-like boundary conditions to model boundaries may be used, by the closure of the domain through actual surface patches. The original approach, by Campbell \cite{Campbell89}, was further developed in Refs. \cite{DeLeffe09,Kulasegaram04}. However, the first consistent formulation is found in Ref. \cite{Ferrand13}, even though some aspects of the consistency of such a method were first analyzed in Ref. \cite{Macia12}. The boundary integrals formulation of Ref. \cite{Ferrand13} considers an analytical kernel factor close to the boundaries, modeled as elements with associated areas and normal vectors.

Later on, a purely numerical boundary integrals methodology was developed in Ref. \citep{Cercos-Pita15b}, which may be easily extended to 3-D simulations. To this end, the boundary is again discretized into elements which represent an actual area and have a normal associated to the wall they are modeling.

The boundary integrals methodology, though still in development, is a powerful choice --- especially for complex geometries and 3-D applications, where the ghost particle approach becomes too complex. Nevertheless, the main setback in Ferrand's formulation \cite{Ferrand13} is the high computational cost introduced by the connectivity between surface elements. There is therefore a need for a faster boundary integrals methodology. On the other hand, the numerical boundary integrals methodology is not as accurate as the semi-analytical formulation, as a truncation error is assumed and the symmetry of the operators is broken due to Shepard renormalization. Effort is hence needed in this regard.

The aim of this work is to improve the accuracy of the numerical boundary integrals formulation, while avoiding a high computational cost. In order to achieve this, an alternative formulation of the Shepard renormalization factor for the boundary integrals methodology, close to boundaries, will be introduced in Section \ref{sec: Boundary Integrals}. Afterwards, an analysis to properly assess the new formulation will be carried out in Section \ref{subsec:gammaEvaluation}. Finally, the new formulation will be tested against different geometries in Section \ref{sec:applications}, where results for 2-D case and a 3-D dam break will be presented.

\section{The role of the Shepard renormalization factor in the Boundary Integrals formulation}
\label{sec: Boundary Integrals} 

\begin{figure}[t]
\begin{center}
\includegraphics[width=0.8\columnwidth]{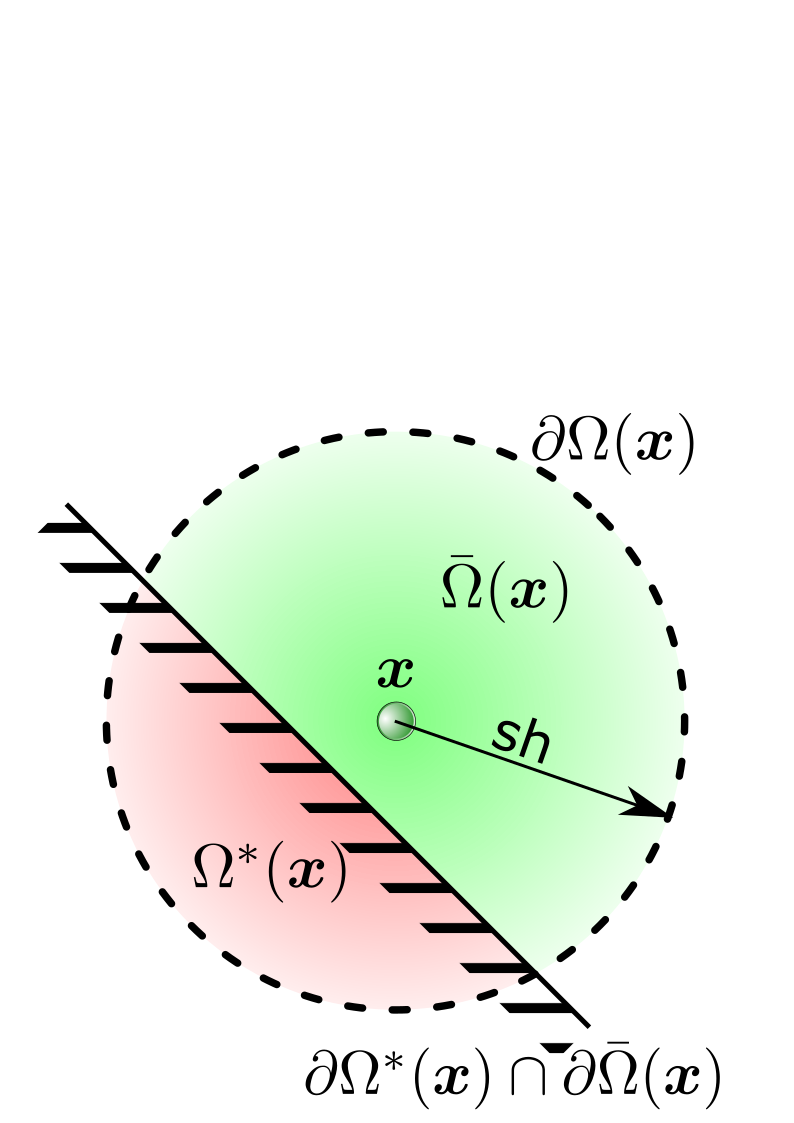}
\end{center}
\caption{Schematic view of a point convolution in the presence of a boundary.}
\label{fig:continuum-boundary}
\end{figure}

Figure \ref{fig:continuum-boundary} shows the schematic view of a point convolution $\bm{x}$ in the presence of a boundary. In such situation, the compact support of the kernel, $\Omega(\bm{x})$, can be divided into two different subdomains. The fluid subdomain is denoted by $\bar{\Omega}(\bm{x})$ and the subdomain across the boundary, $\Omega^{*}(\bm{x})$. For the sake of simplicity, from now on, kernel function notation will be $\Omega$, $\bar{\Omega}$ and $\Omega^*$ respectively, following the notation of Ref. \cite{Antuono10}.

Although fluid information in $\bar{\Omega}$ is always known, information beyond the wall ($\Omega^{*}$) is not known \textit{a priori}, hence different techniques, as the ones mentioned in Section \ref{1. Introduction}, have been developed in this regard. In this work, we will focus on the gradient operator, which plays an important role in SPH simulations. Expanding the continuous convolution of the first order differential operator, for a generic field $f$ and considering both subdomains, yields:

\begin{equation}
  \begin{split}
\label{eq:subdomainInt}
\sph{\mathcal{D} f(\bm{x})} & = 
	\int_{\bar{\Omega}}
	f(\bm{y}) \nabla W_h (\bm{y} - \bm{x})\ d\bm{y}  \\
        &
	+ \int_{\Omega^{*}} f(\bm{y})
	\nabla W_h (\bm{y} - \bm{x})\ d\bm{y} ,
  \end{split}
\end{equation}
where the bracket operator $\sph{\ }$ represents SPH interpolation, as in Ref. \cite{DeLeffe09}, and $\bm{y}$ is a generic point inside the kernel support. The kernel $W_h (\bm{y} - \bm{x})$ is supposed to depend only on the distance between points, divided by a characteristic constant length $h$. The kernel has compact support, i.e. it vanishes for $\vert \bm{y} - \bm{x} \vert > s \ h$, where $s$ is an integer greater than $0$. Applying the divergence theorem, and the kernel properties, a popular expression for the boundary integrals formulation \cite{Ferrand13} is found:
\begin{equation}
  \begin{split}
    \label{eq:finalDiffOperator}
\sph{\mathcal{D} f(\bm{x})} = 
& \frac{1}{\gamma(\bm{x})}
\left(
	\int_{\bar{\Omega}} 
	f(\bm{y}) \nabla W_h (\bm{y} - \bm{x})\ d\bm{y}  \right.\\
        &
        \left.
			+ \int_{\partial \bar{\Omega}}
				f(\bm{y}) \cdot \bm{n}(\bm{y}) W_h(\bm{y} - \bm{x})\ d\bm{y} 
				\right)
\end{split}
\end{equation}
where the Shepard renormalization factor $\gamma(\bm{x})$ is introduced in order to keep consistency. It is defined, in continuous form, as:

\begin{equation}
\label{eq:ShepardFactor}
\gamma(\bm{x}) := 
	\int_{\bar{\Omega}} W_h (\bm{x} - \bm{y})\ d\bm{y} .
\end{equation}

Or as the discretized version:

\begin{equation}
\label{eq:discreteShepardFactor}
\gamma (\bm{x}_i) = \dsty{ \sum_{j \in Fluid}} W_h (\bm{x}_j - \bm{x}_i) V_j ,
\end{equation}
where $\bm{x}_i$ and $\bm{x}_j$ are the positions of a generic particle and its neighbors, respectively, and $V_j$ their volumes.

Regarding the applicability of the boundary integrals formulation, the main setback comes from the fact that the Shepard renormalization factor, $\gamma (\bm{x})$, breaks the symmetry form of the operators, and therefore momentum and energy conservation cannot be assured anymore. This is because, at the boundary, $\gamma(\bm{x}) \neq \gamma(\bm{y})$ when $\bm{x} \neq \bm{y}$.
Additionally, in multiphase flows, the Shepard renormalization factor integral domain near the free surface boundary should consider all fluid domains in order to be consistent with boundary integrals formulation. 
However, it is common practice to only model the heavier phase, which is generally the most relevant. This reduces computational effort and has indeed become one of the greatest advantages of SPH for free surface flow simulations. On the other hand, the domain at the free surface is then incomplete, which causes a wrong computation of the Shepard renormalization factor.
In order to avoid this issue, the correct procedure, as argued in Ref. \cite{Colagrossi09}, would be to explicitly model the lighter phase.

Another problem of the methodology occurs when the boundary integrals formulation moves on to the discrete level, i.e. when Eq. \eqref{eq:discreteShepardFactor} is applied. In this context, even far from the boundary, $\gamma(\bm{x_i}) \neq \gamma(\bm{y_j})$ when $\bm{x_i} \neq \bm{y_j}$, which may lead to values of the Shepard renormalization factor bigger than one. It might also happen that tensile instabilities appear because of the nature of the SPH interpolating kernels close to boundaries. Issues derived from these instabilities, such as sensitivity to particle disorder, clamping, or clustering clearly affect the computation of the Shepard factor, and might end up in unacceptable Shepard values.

Given all these facts, this work aims to develop a new boundary integrals formulation that overcomes these disadvantages by defining a new integral at the boundaries that allows to transform the whole volume integral into a surface boundary integral.

\subsection{Alternative geometrical formulation of the Shepard renormalization factor}
\label{subsec:newKernel}

A methodology to transform the volume integral to compute the Shepard renormalization factor (\ref{eq:discreteShepardFactor}) into a surface integral was described in Ref. \cite{feldman2007}, extended later as a semi-analytical method for a generic 2-D profile in Ref. \cite{Leroy14} and finally incorporated in Ref. \cite{violeau2014_spheric} to 3-D applications.
However, both approaches can be unified as described in this Section.

Notice that the kernel is normalized, and we may integrate it by splitting its support as
\begin{equation}
\label{eq:newKernelJust}
\int_{\bar{\Omega}}
	W_h (\bm{y} - \bm{x})\ d\bm{y} \,
	+ \int_{\Omega^{*}} W_h (\bm{y} - \bm{x})\ d\bm{y} = 1 .
\end{equation}

From the definition of the Shepard factor, Eq. \eqref{eq:ShepardFactor}:
\begin{equation}
  \label{eq:OmegaShepard}
  \gamma(\bm{x}) + 
  \int_{\Omega^{*}} W_h (\bm{y} - \bm{x})\ d\bm{y} =
  1 .
\end{equation}

Our goal is to express the Shepard renormalization factor purely as a surface integral. If we can find a function $F(\bm{y} - \bm{x})$ such that
\begin{equation}
\label{eq:newKernelCondition}
\int_{\Omega^{*}} \nabla \cdot
\left[
(\bm{y} - \bm{x})\ F (\bm{y} - \bm{x}) 
\right] = 
\int_{\Omega^{*}} W_h (\bm{y} - \bm{x})\ d\bm{y} ,
\end{equation}
then, $\gamma (\bm{x})$ may be obtained from a boundary integral through the application of the divergence theorem:
\begin{equation}
\label{eq:ShepardRenormalizationNew}
\gamma(\bm{x}) =
         1 + \int_{\partial \bar{\Omega}}
         \bm{n} (\bm{y}) \cdot
         (\bm{y} - \bm{x})\ F(\bm{y} - \bm{x})
         \ d\bm{y} .
\end{equation}

We may therefore formulate $\gamma (\bm{x})$ in terms of surface integrals, which are purely geometrical, unlike volume integrals, which may change in time since the particles will move. They will also be clearly independent on any free surfaces that may appear, whilst volume integrals would be affected by the integration over an area with no particles.

Since the definition of Eq. \eqref{eq:ShepardRenormalizationNew} must apply for every integral, the general solution is given by the equality of the integrands.
Moreover, as the kernel $W_h$ depends only on $ \rho := \rvert \bm{y} - \bm{x} \rvert $, the distance between $\bm{x}$ and $\bm{y}$, we may seek a $F$ that also depends only on $\rho$, which leads to
\begin{equation}
\label{eq:kernelExpression}
\frac{1}{\rho^{d-1}} 
\frac{d \left( \rho^d \, F (\rho) \right) }{d \rho} = 
	W_h (\rho) ,
\end{equation}
where $d$ is the spatial dimension.

The solution to Eq. \eqref{eq:kernelExpression} depends, of course, on the choice of kernel. For the Wendland kernel \cite{Wendland95}, and $s=2$, expressions for $F(\rho)$ in reduced (non-dimensional) form are given by:
\begin{equation}
\label{eq:kernelSolution1}
F(\rho) = 
  \frac{1}{h^d} \ \tilde{F} \left (\frac\rho{h} \right) ,
\end{equation}
with the function of the reduced distance $q := \rho / h$, expressed in 2-D as
\begin{align}
\label{eq:kernelSolution2D}
\notag
\tilde{F} (q) = 
    &
  - \frac{7}{64 \pi}
  \left[
    \frac{2}{7} q^5 -
    \frac{5}{2} q^4 +
    8 q^3 -
    10 q^2 +
    8
    \right] \\
  & \qquad - \frac{1}{2 \pi \, q^2} ,
\end{align}
and in 3-D:
\begin{align}
\label{eq:kernelSolution3D}
  \notag
\tilde{F} (q) = 
    &
  - \frac{21}{256 \pi} 
  \left[
    \frac{1}{4} q^5 -
    \frac{15}{7} q^4 +
    \frac{20}{3} q^3 -
    8 q^2 +
    \frac{16}{13}
    \right] \\
  & \qquad - \frac{1}{4 \pi \, q^3} .
\end{align}

Both expressions in Eqs. (\ref{eq:kernelSolution2D}) and (\ref{eq:kernelSolution3D}) can in fact be separated into a polynomial part and a divergent part, that we will call from now on $\tilde{F}_{P}$ and $\tilde{F}_{D}$ respectively. Therefore, in order to be consistent with the notation $F(\rho) = F_{P} (\rho) + F_{D} (\rho)$.
This implies a set of restraints and a different treatment of the kernel function that will be discussed in next section.

\subsection{Efficient evaluation of $\gamma$}
\label{subsec:gammaEvaluation}

\begin{figure}[t]
\centering
    \includegraphics[width=0.9\columnwidth]{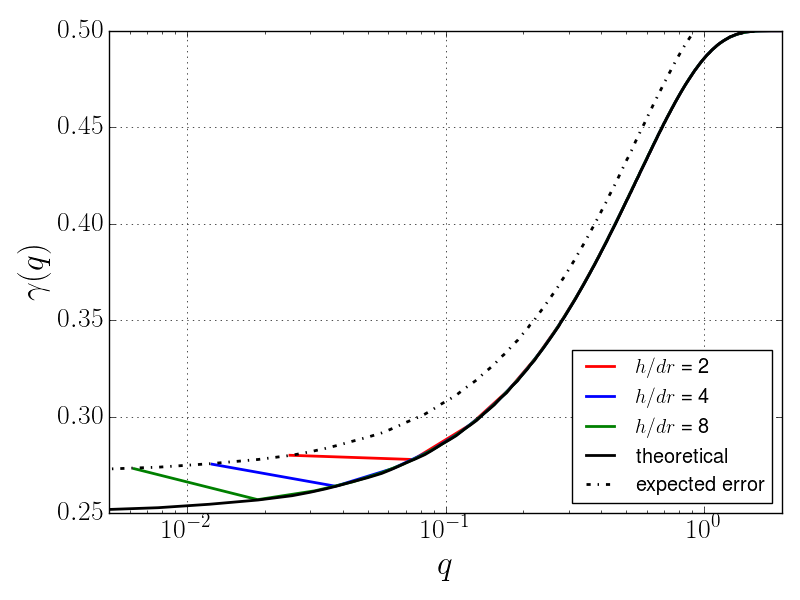}
    \caption{Shepard value at the contour near a 90$^\circ$ corner for different kernel supports computed with $F(q)$ and compared to the analytical value. Dashed line is the black solid line displaced upwards $C = 0.021$, and represents the expected error. Horizontal axis: normalized distance $q$ to the corner, vertical axis: Shepard value $\gamma(\bm{x})$.
      \label{fig:corner}}
\end{figure}

The effect of having a singular kernel -- such as $F(q)$ -- is especially significant near a boundary, where it has a greatest influence on the particle interpolation. Nonetheless, there exist several alternatives to deal with this fact. From all the options that could be considered, increasing the resolution might be the more straight-forward approach. However, results are still not assured, while moving on to an increase above 50 times the fluid particle resolution results in an unacceptable increase of the computational cost, keeping in mind that a fast and reasonably accurate solution is being sought. To show it, Figure \ref{fig:corner} illustrates how a resolution increase affects the Shepard renormalization factor $\gamma(\bm{x})$ compared to the theoretical solution 
for a fluid particle approaching a 90$^\circ$ corner along one of the boundaries. The divergent term of the kernel has an influence on the last particle interpolation, giving a wrong value of the Shepard renormalization factor which is independent of the resolution. The dashed line in Figure \ref{fig:corner} represents the expected error derived from this last particle interpolation. As it can be appreciated, the error is nearly constant, in this case with a value of about $C = 0.021$. Since the value is the same for different resolutions, it can be guessed that the error follows the Shepard curvature.

Alternatively, Eq. \eqref{eq:ShepardRenormalizationNew} could be solved analytically.
Such approach has been already considered in \cite{violeau2014_spheric}, leading however
to involved expressions that, again, would be too costly from the computational point of view. 
Finally, another possibility, based on the previous one, would consist on solving once for multiple situations and tabulate the results. Unfortunately, a very high resolution would be needed again to obtain an accurate enough solution.
It seems that none of these options are able to improve the solution. There is, however, another one, to be analyzed. 
Re-writing the expression in Eq. (\ref{eq:ShepardRenormalizationNew}) for a boundary that is split into different $S_j$ patches yields:
\begin{equation}
\label{eq:generalShepardSingular2}
\gamma(\bm{x}) = 
	1 + \sum_j \int_{S_j} \bm{n} (\bm{y}) \cdot (\bm{y} - \bm{x}) \ F ( \bm{y} - \bm{x} )\, d\bm{y} .
\end{equation}

\begin{figure}[t]
\centering
    \includegraphics[width=0.9\columnwidth]{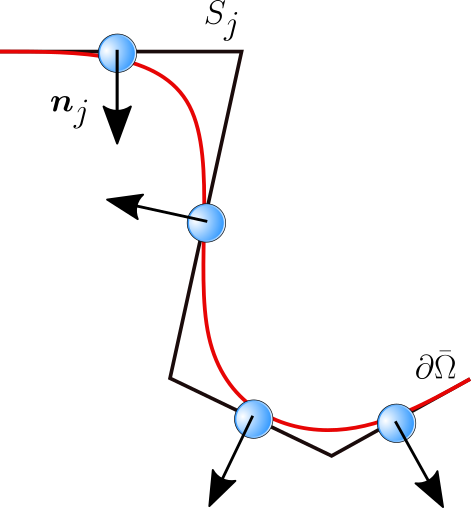}
  \caption{Discretization of the continuous boundary $\partial \bar{\Omega}$ (red line) to a set of discretized planar patches $S_j$ (black line)
    \label{fig:boundary-discretization}}
\end{figure}

If the curvature of the patches is neglected, they may be approximated by flat segments, so that normals are constant, as depicted in Figure \ref{fig:boundary-discretization}. In this case,
\begin{equation}
\label{eq:generalShepardSingular3}
\gamma(\bm{x}) - 1 \simeq 
	\sum_j \bm{n}_j \cdot (\bm{y}_j - \bm{x} )
		\int_{S_j}    F (\bm{y} - \bm{x}) \, d\bm{y} .
\end{equation}

The expression $\bm{n}_j \cdot (\bm{y}_j - \bm{x} )$, with $\bm{y}_j\in S_j$, is taken out of the integral since, for a flat patch, it equals $r_j$, the distance between $\bm{x}$ and the patch.
The expression above is in fact similar to the ones proposed in Ref. \cite{Ferrand13}.

Finally, going back to Eqs. (\ref{eq:kernelSolution2D},~\ref{eq:kernelSolution3D}), it has already been noticed that the kernel can be decomposed into a polynomial term,  $F_\text{P}$,  and a divergent term  $F_\text{D}$.

The integral of the polynomial term will still be solved numerically within the traditional boundary integrals formulation ---either the semi-analytical \cite{Ferrand13} or the purely numerical \citep{Cercos-Pita15b} one, whereas the divergent term is solved analytically. In \ref{sec:appendix} it is shown that, based on the previous approximations, one can get in 2-D to:

\begin{equation}
\label{eq:subtended_angle2D}
\bm{n}_j \cdot (\bm{y}_j - \bm{x} )
		\int_{S_j}   F_\text{D} (\bm{y} - \bm{x}) \, d\bm{y}
                = 
		\displaystyle
                  -\frac1{2\pi} \Delta\theta_j ,
\end{equation}

and in 3-D to:

\begin{equation}
\label{eq:subtended_angle3D}
\bm{n}_j \cdot (\bm{y}_j - \bm{x} )
		\int_{S_j}   F_\text{D} (\bm{y} - \bm{x}) \, d\bm{y}
                = 
		\displaystyle
                  -\frac1{4\pi} \Delta\Omega_j .
\end{equation}

In 2-D, $\Delta\theta_j$ is the angle subtended by segment $S_j$ at point $\bm{x}$, which can be easily computed. Similarly, in 3-D $\Delta\Omega_j$ is the solid angle subtended by patch $S_j$  at point $\bm{x}$. 

The approach described in \cite{Mayrhofer15b}, where Eq. (\ref{eq:generalShepardSingular3}) is directly addressed, yields a complex and costly formulation.
That is partially solved in \cite{violeau2014_spheric}, where the same expression (\ref{eq:generalShepardSingular3}) is addressed, applying this time the Gauss' theorem to transform it in a line integral, which fits the semi-analytical formulation described in \cite{Ferrand13}, but has a poor performance in purely numerical approaches \cite{Cercos-Pita15b}.
In \ref{sec:appendix2} efficient ways to evaluate the subtended angle from Eqs. (\ref{eq:subtended_angle2D},~\ref{eq:subtended_angle3D}), both in semi-analytical and purely numerical contexts, are described.
\added[id=rev2]{
Incidentally, we may emphasize that different expressions can be found, depending on the arbitrary discretization patch shape selected, in a similar fashion of mesh-based methods \citep{Sozer14}.
In these  algorithms it is assumed that the boundary is discretized in straight line segments for 2-D applications, while for 3-D applications different discretisations have been considered: triangles for the semi-analytical approach, and squares for the purely numerical one. Indeed, in the semi-analytical approach the computational overhead of the connectivities is generally accepted, taking advantage of the well-known surface triangulation properties \citep{Delaunay34}. Conversely, in the purely numerical approach the boundary is discretized in square patches, as a natural 2-D extension of the volumetric discretization usually applied in SPH.}
Comparing the computational performance and results quality of both approaches is out of the scope of this work. Nevertheless, an attempt to compare the performance of the new formulation and the traditional approach within the methodology of numerical boundary integrals is carried out in Section \ref{subsec:applications:dam-break}.

It is noteworthy that Eq. \eqref{eq:subtended_angle2D} in 2-D correctly predicts that for a point, $\bm{x}$, on one of the boundary patches,
\begin{equation}
  \label{eq:limKernel}
  \gamma(\bm{x}) - 1  \rightarrow
  -\frac1{2\pi} \pi = -\frac12 ,
\end{equation}
as long as only the patch containing $x$ contributes to the summation (because all the others are either far away, or aligned with the patch). This is the expected result: $\gamma(\bm{x})=1/2$ close to a flat boundary.
The same holds true in 3-D applying Eq. (\ref{eq:subtended_angle3D}).

\begin{figure}[t]
\begin{center}
\includegraphics[width=0.9\columnwidth]{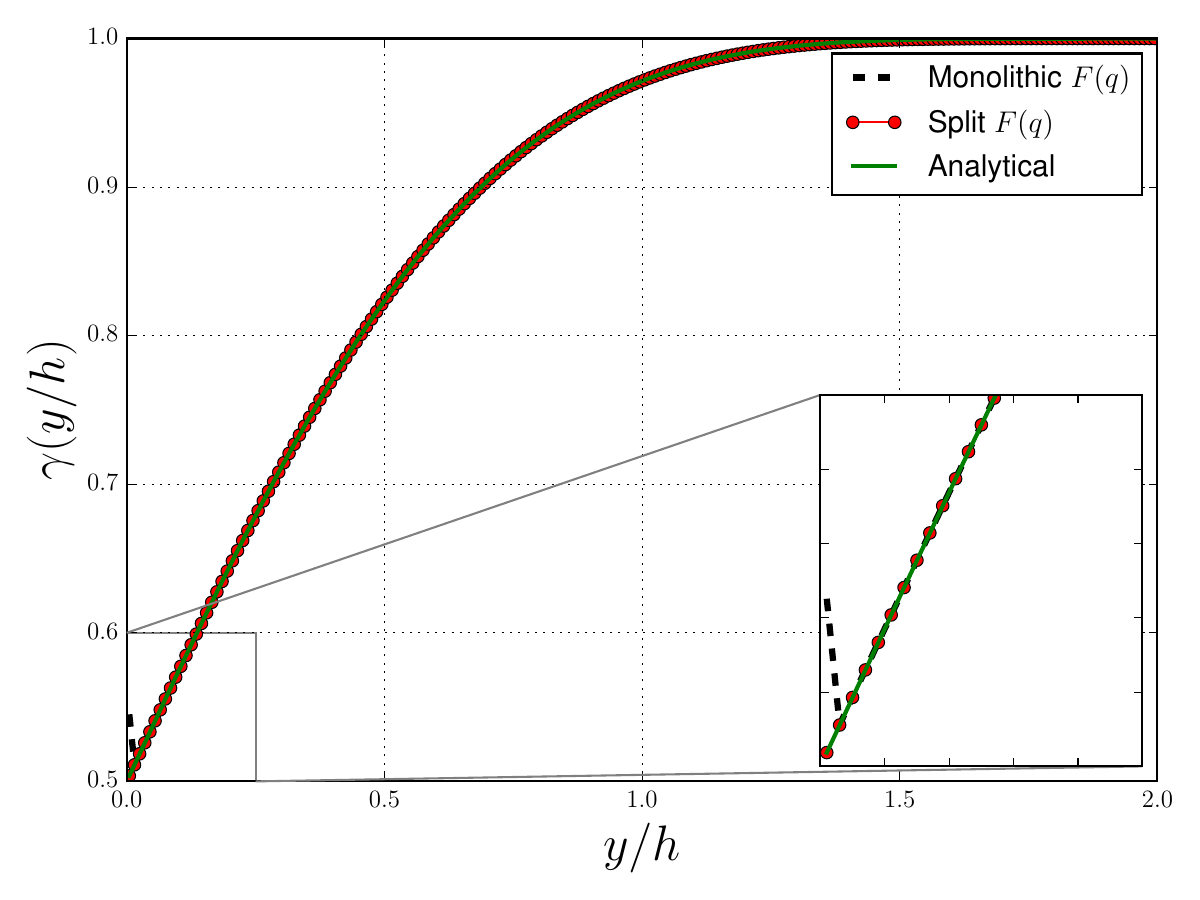}
\end{center}
\caption{Shepard value for a fluid particle approaching a straight boundary. Black dashed line: monolithic expression from Eq (\ref{eq:kernelSolution2D}), red line: the same expression, but with the semi-analytic formulation as in Eq. (\ref{eq:subtended_angle2D}), green line: analytical value of the Shepard renormalization factor.
\label{fig:straight}}
\end{figure}

Therefore, if a point is detected at the contour, then it is just necessary to add $- \ahalf$ to the integration, such that the point itself will be ignored in the numerical integration.
Performance of Eq. \eqref{eq:generalShepardSingular3} has been tested and the result is shown in Figure \ref{fig:straight}, which shows the Shepard renormalization factor value for a fluid particle approaching a horizontal wall, from both the monolithic expression of the new kernel ($F(\bm{q})$) ---i.e. the straight forward approach---, and from expressions \eqref{eq:subtended_angle2D} and \eqref{eq:subtended_angle3D} ---in which the divergent part of the integral is solved analytically---.
Both are seen to perform well far enough of the wall, whilst their behavior starts to differ when they are very close to the wall.
Here is where the divergent part weights more and therefore, having an exact expression for this part makes the solution to be accurate near the wall.
Otherwise, the increment in the particle resolution would need to be huge to diminish the error near the wall.

\begin{figure*}[t]
\centering
\includegraphics[width=0.45\textwidth]{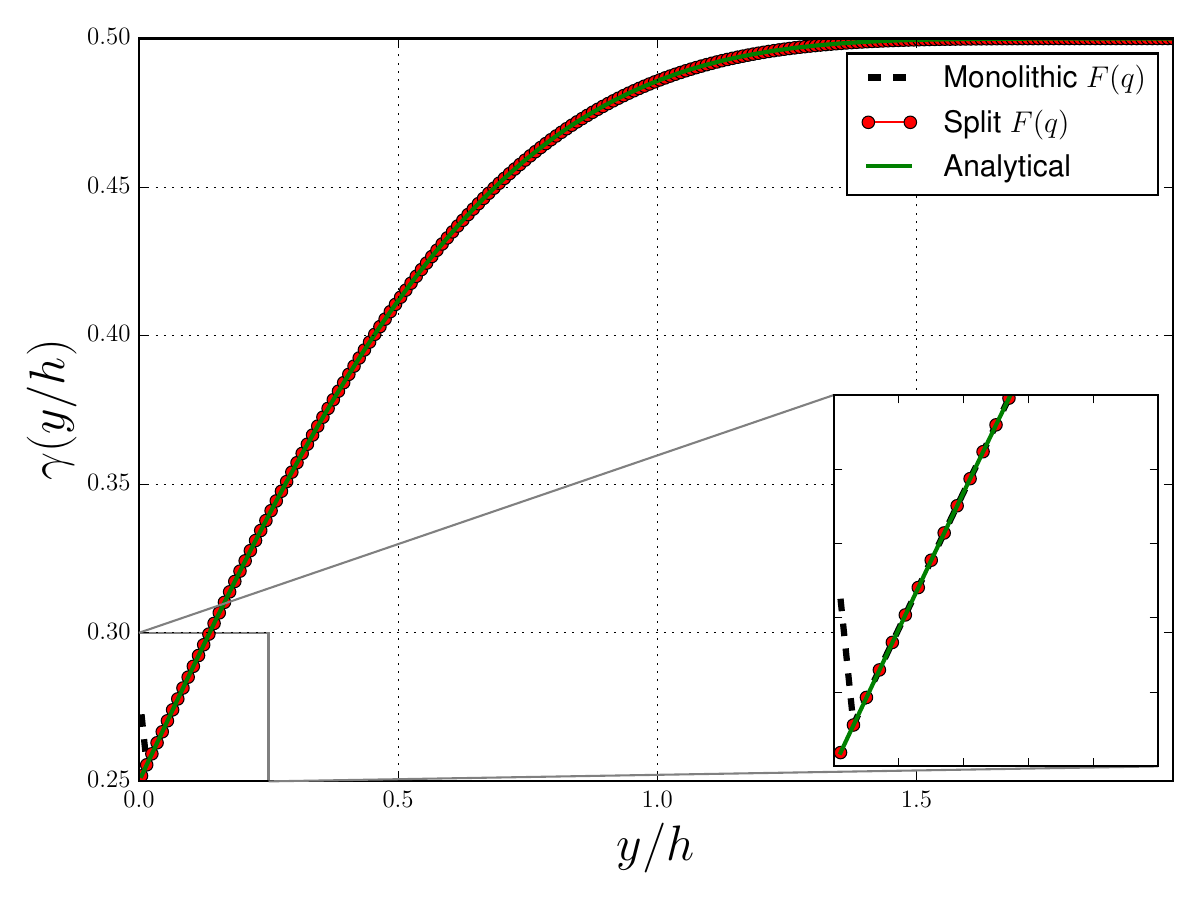}
\includegraphics[width=0.45\textwidth]{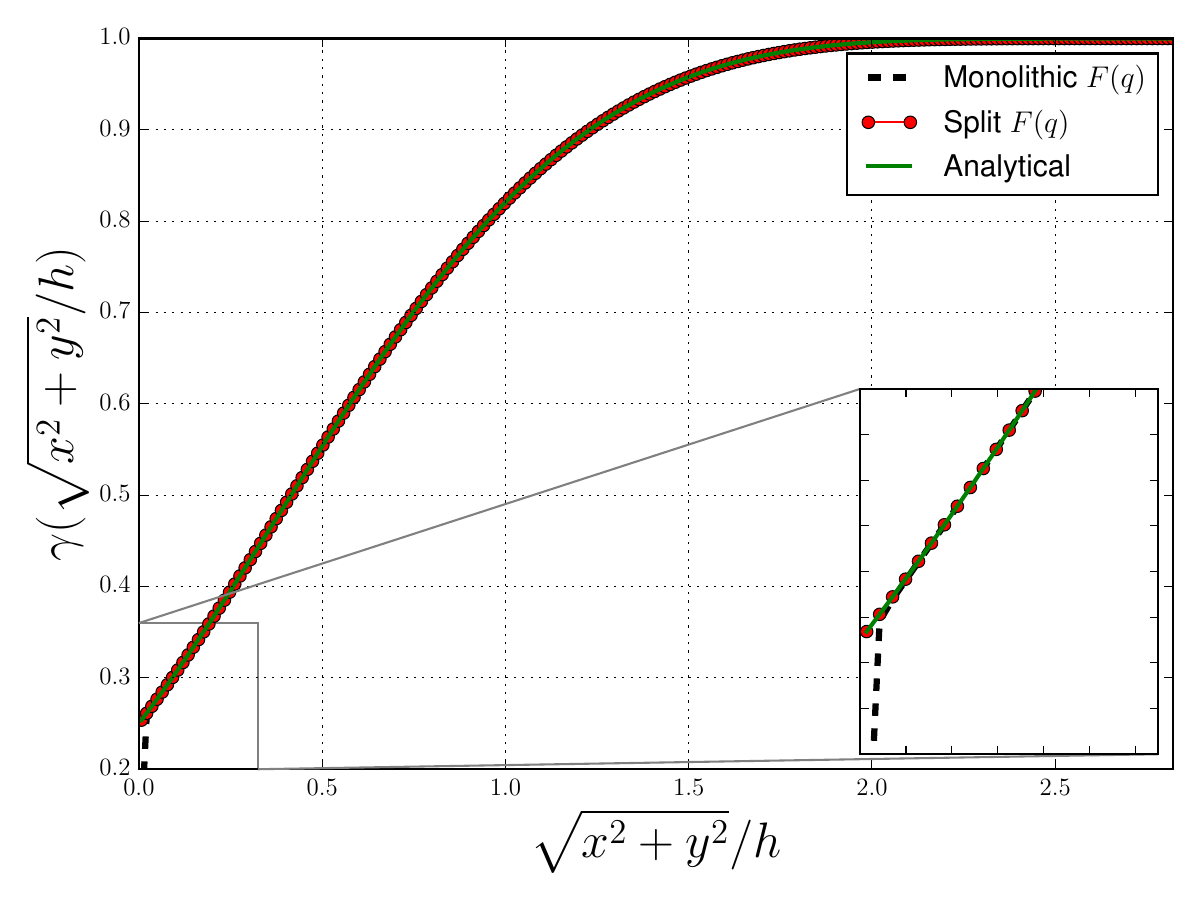}
\caption{Shepard value for a 90$^\circ$ corner, for approaches along the vertical wall (left) and the diagonal (right). Legend as in Figure \ref{fig:straight}
  \label{fig:90corner}}
\end{figure*}

This derivation is extensible to any kind of planar boundary, including any angled corner.
Figure \ref{fig:90corner} shows the results for a 90$^\circ$ corner. Performance is similar to the one shown in Figure \ref{fig:straight}, where a straight boundary is considered.
In fact, results in Figure \ref{fig:90corner} (left) are equivalent to the test results shown in Figure \ref{fig:corner}.
It can be appreciated that the correct value of $\gamma(\bm{x}) = 0.25$ is now reached at the corner when the divergent part of the kernel is integrated exactly with the new formulation.

\begin{figure*}[t]
\centering
\includegraphics[width=0.9\columnwidth]{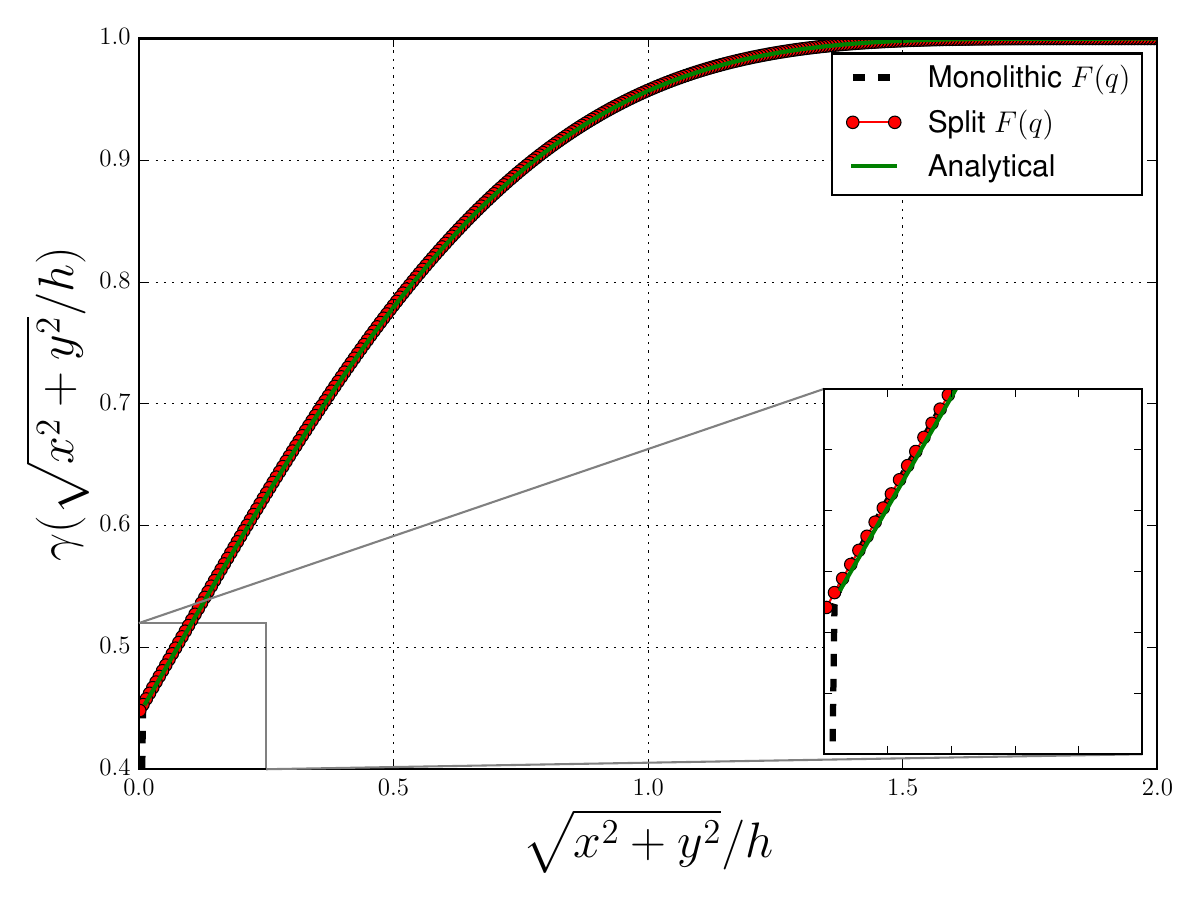}
\caption{Shepard value for a circular boundary for a particle approaching along the radius of the circle from the center to the boundary. Legend as in Figure \ref{fig:straight}
  \label{fig:rounded}}
\end{figure*}

The new formulation is not only suitable for planar boundaries but for general boundaries indeed. Figure \ref{fig:rounded} represents the Shepard value at a circular boundary for a particle approaching radially from the center of the domain. Performance is seen to be similar to the previous tests shown. The new formulation predicts correctly the value when approaching the boundary even having considered flat segments. Again, the monolithic formulation is not able to capture the correct value of the Shepard renormalization factor close to the boundary due to the effect of the singularity in the kernel.

\begin{figure}[t]
\begin{center}
\includegraphics[width=0.8\columnwidth]{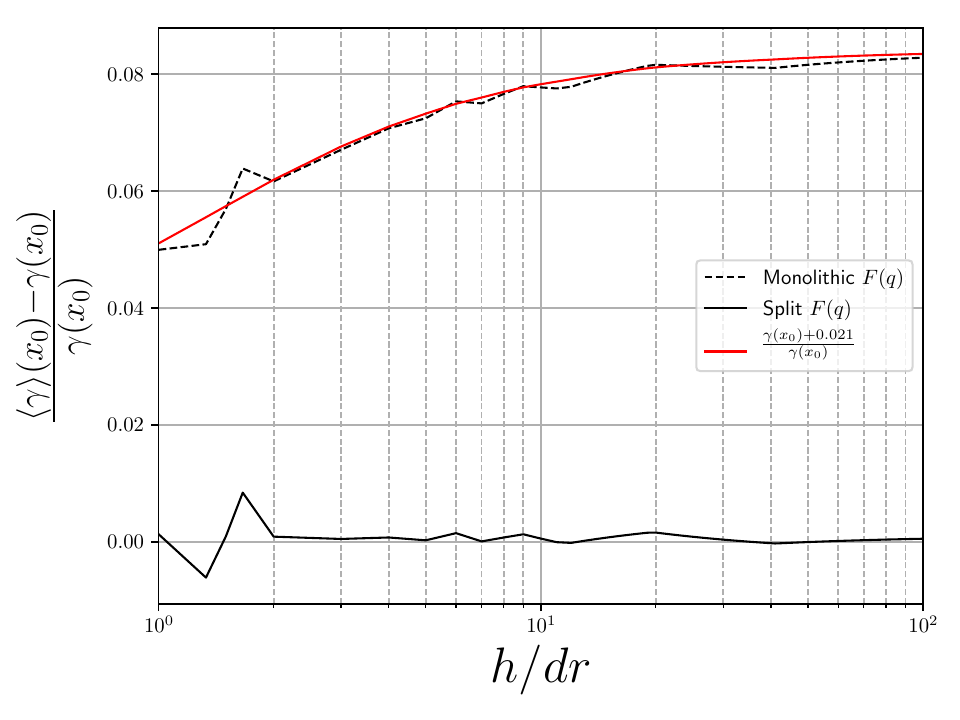}
\end{center}
\caption{Relative error in the Shepard value at fixed point, close to a 90$^\circ$ corner as a function of $h/(\Delta r)$. Dashed line: monolithic expression; red line: approximation to the former; solid line: split expression.
  \label{error_vs_hdr} }
\end{figure}

\added[id=rev2]{In order to assess the accuracy of the method as a
  function of the $h/(\Delta r)$ ratio, which controls the number of
  neighbors in an SPH calculation, we select the point closest to the
  corner of Figure \ref{fig:90corner} (the one with the worst
  results), and evaluate the relative error in the Shepard value, as
  this ratio is decreased. In Figure \ref{error_vs_hdr} the monolithic
  expression is seen to improve slightly as fewer neighbors are
  considered. Meanwhile, the semianalyitic method is seen to result in
  much lower errors. Both methods are affected by fluctuations as the
  number of neighbors becomes low. It is obvious that these are caused
  by the part that is common to both methods: the polynomial. Also,
  the monolithic part is well fit by the expression with a $C = 0.021$
  displacement mentioned above and in Figure \ref{fig:corner}.}

\section{Applications}
\label{sec:applications}
\subsection{Hydrostatic 2-D case}
\label{subsec:applications:hydrostatic}
With the results shown in Section \ref{subsec:gammaEvaluation}, it is expected that some current drawbacks of the boundary integrals formulation are avoided and results improve for any kind of geometry that is tested.
In order to carry out a more relevant test, a still liquid test case has been carried out with the open-source free tool AQUAgpusph \cite{Cercos-Pita15b}, with both the standard formulation and the new formulation. 

A weakly-compressible $\delta$-SPH model is solved. The $\delta$-SPH scheme \cite{Antuono12, Antuono15, Cercos-Pita16a} is considered in order to avoid numerical instabilities. The standard Navier-Stokes continuity and momentum equations for a barotropic fluid are solved including diffusive terms for both equations. See for instance Refs. \cite{Antuono12, Cercos-Pita17}. A stiffened linear Equation of State, as in Refs. \cite{Antuono10, Cercos-Pita16a} is used.

\begin{figure}[t]
\begin{center}
\includegraphics[width=0.9\columnwidth]{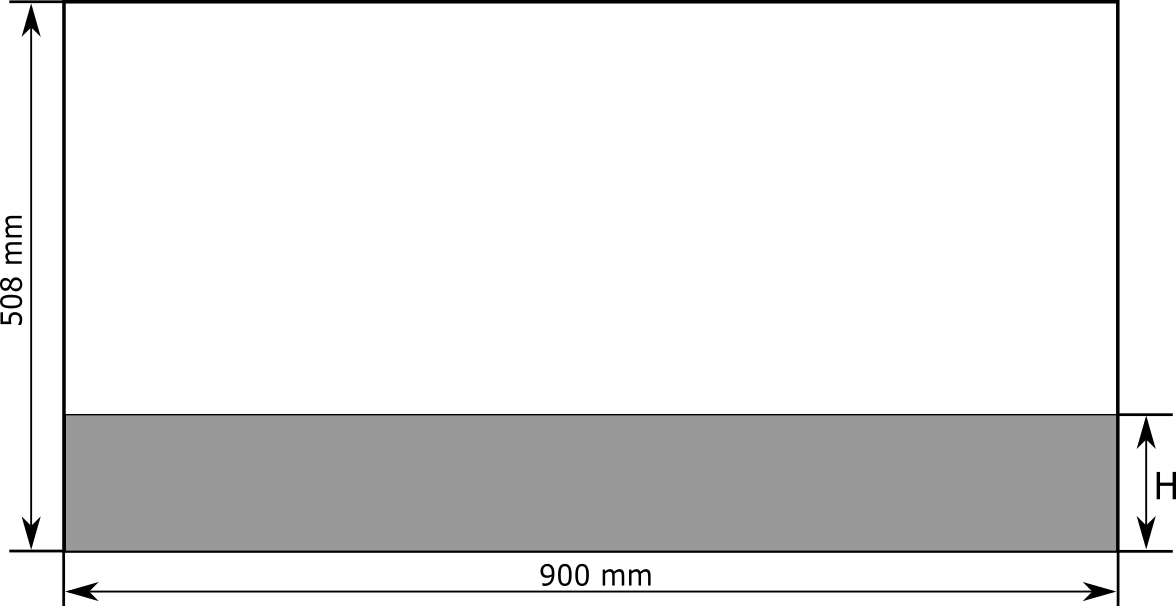}
\end{center}
\caption{Geometry of the 2-D hydrostatic tank.}
\label{fig:Geometry}
\end{figure}

The case consists of a tank set at rest for a certain period of time. The geometry is shown in Figure \ref{fig:Geometry}. 
The aim is to assess the performance of the new formulation introduced in previous section, to check whether the new boundary term properly computes the integrals.
This is a relatively simple test that nevertheless allows an easy assessment of the features that should be improved. In particular, errors in the consistency of the operators, which may lead to spurious velocities and nonphysical pressure values, and to the wrong computation of forces in the tank.

Moreover, problems mainly associated to the escape of particles from the boundary domain, which can occur with the current formulation, are expected to be avoided.
In addition to that, the traditional formulation makes it impossible to properly correct terms involving the divergence of the velocity, due to inconsistencies that arise in the presence of a free surface. This issue is now fixed, since with the new formulation the divergence of the velocity term is appropriately renormalised with the Shepard factor term.
         
Density is set as $\rho = 998 \mathrm{kg / m^3}$ and dynamic viscosity $\mu$  as the standard value $\mu = 8.94 \times 10^{-4} \mathrm{Pa \cdot s}$. There is therefore no artificial viscosity in the simulation.
Filling level is chosen as $H = 92 \mathrm{mm}$.
Three different resolutions have been tested, with simulations with $10 000$, $50 000$ and $100 000$ particles. Also, different supports ($h / dr = 2$, $3$ and $4$) have been tried for the finest resolution.
No boundary forces have been implemented to stop fluid particles from penetrating the solid walls.

\begin{figure}[t]
\centering
    \includegraphics[width=0.22\textwidth]{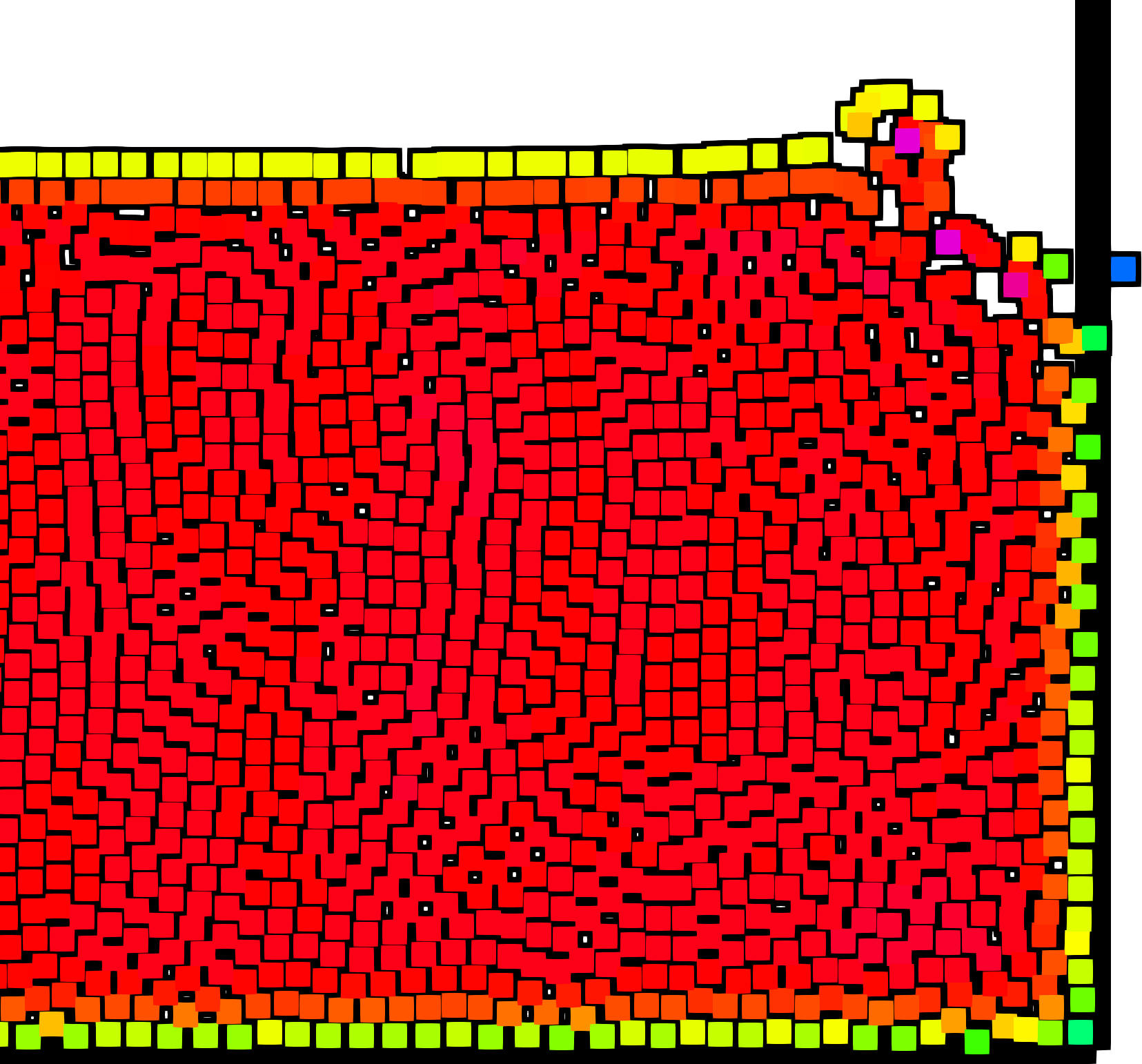}
    \includegraphics[width=0.22\textwidth]{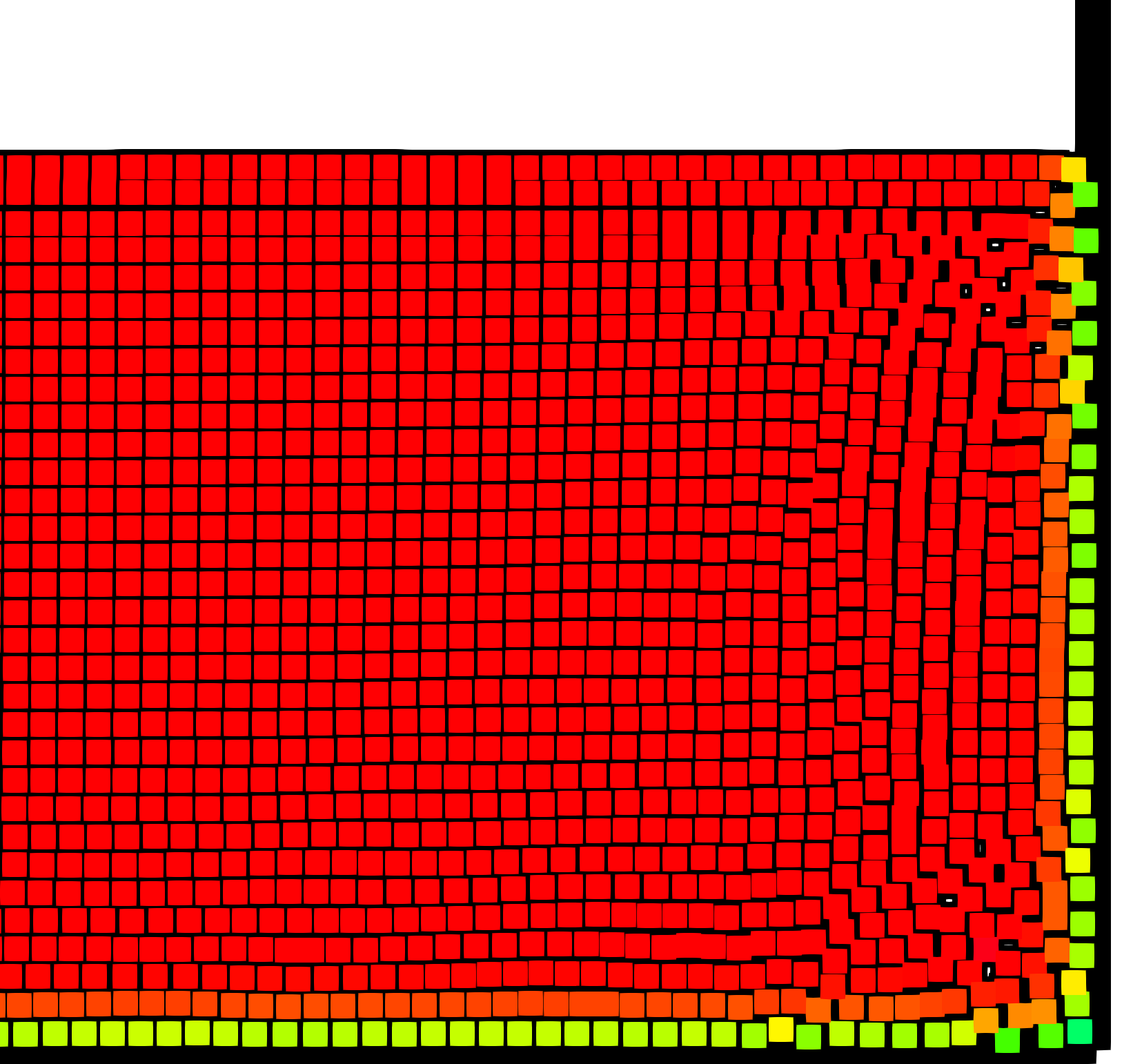} \\
    \includegraphics[width=0.45\textwidth]{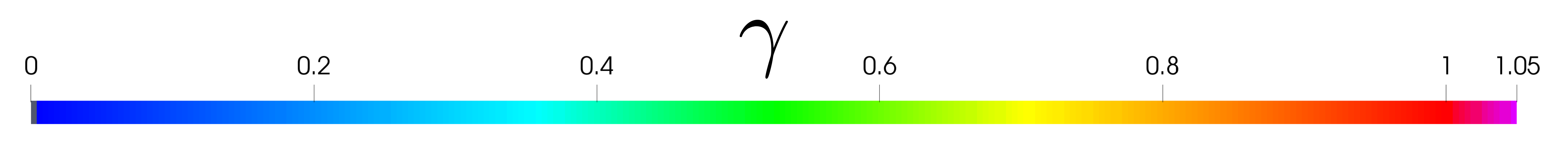}
  \caption{Detail of the Shepard renormalization factor field at the right corner of the Hydrostatic 2-D tank set at rest for the traditional Shepard formulation (left) and the new Shepard geometrical formulation (right). Compact support $h / dr = 2$ and number of particles $N = 10000$. Reduced time is $t \sqrt{g / H} = 9.32$.
    \label{fig:shepardTLD}}
\end{figure}

Figure \ref{fig:shepardTLD} shows the tank at rest for the same simulation time, $t \sqrt{g / H} = 9.32$.
The traditional formulation of the Shepard renormalization factor leads to numerical errors in the interpolation. As can be clearly appreciated, these errors make the particles at the free surface move, with some of them jumping out of the tank. The simulation then turns unstable, even at early stages.
Conversely, the new formulation computes the value of the Shepard renormalization factor exactly at the boundary, and therefore, particles remain without spurious movement and the simulation remains stable in time.

Besides such gross scale instability-related issues, close to the contact line, Shepard renormalization factor errors can be appreciated all along the free surface, when the original expression from Eq. \eqref{eq:discreteShepardFactor} is considered.
Furthermore, it can be appreciated that the original Shepard formulation is prone to produce wrong values, $\gamma(\bm{x}) > 1$, when tensile instabilities occur.

\begin{figure}[t]
\begin{center}
\includegraphics[width=0.9\columnwidth]{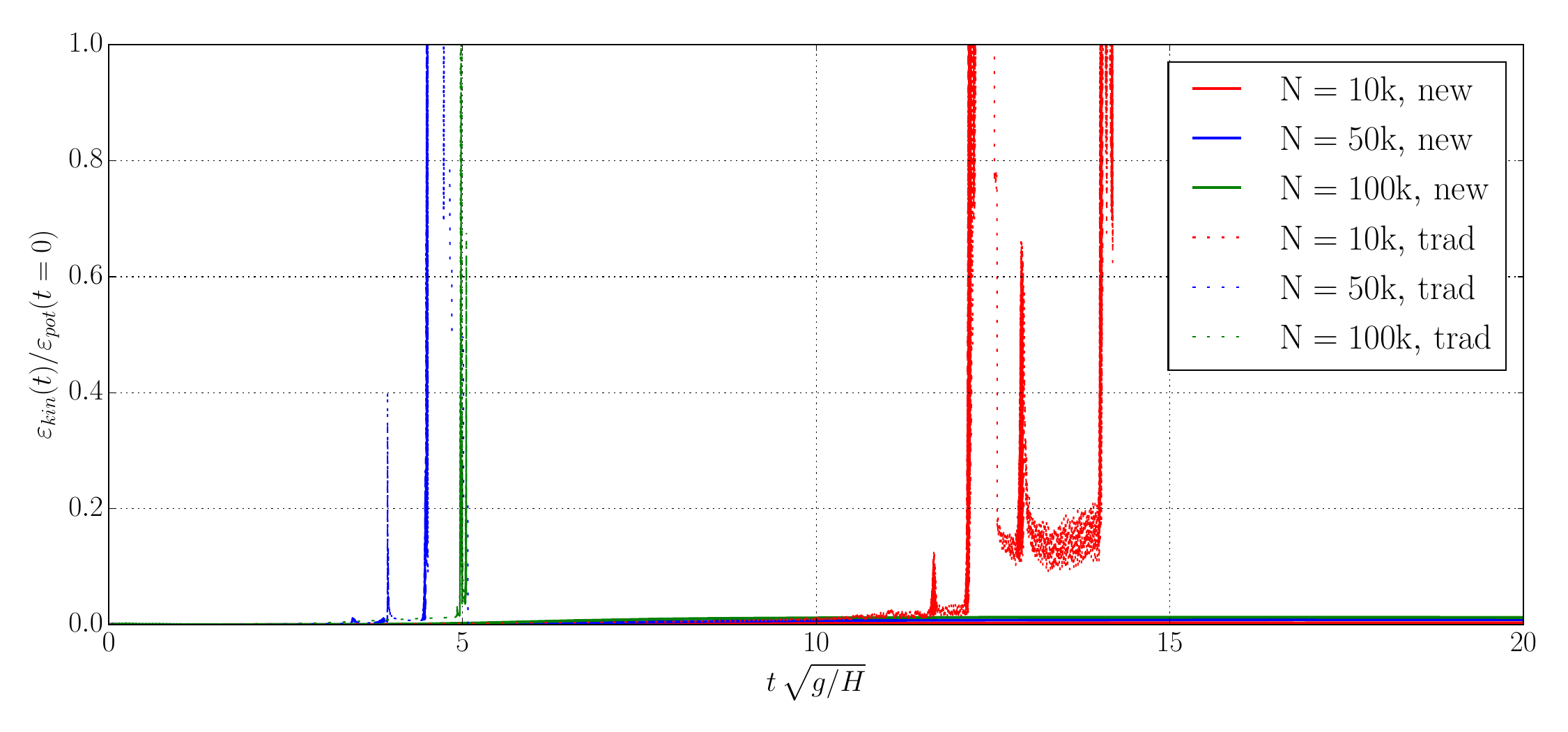}
\end{center}
\caption{Evolution of the reduced kinetic energy for the hydrostatic case. Solid lines: new formulation for three different resolutions ($10 000$, $50 000$ and $100 000$), dashed lines:  traditional formulation.}
\label{fig:kineticEnergy}
\end{figure}

\begin{figure}[t]
\begin{center}
\includegraphics[width=0.9\columnwidth]{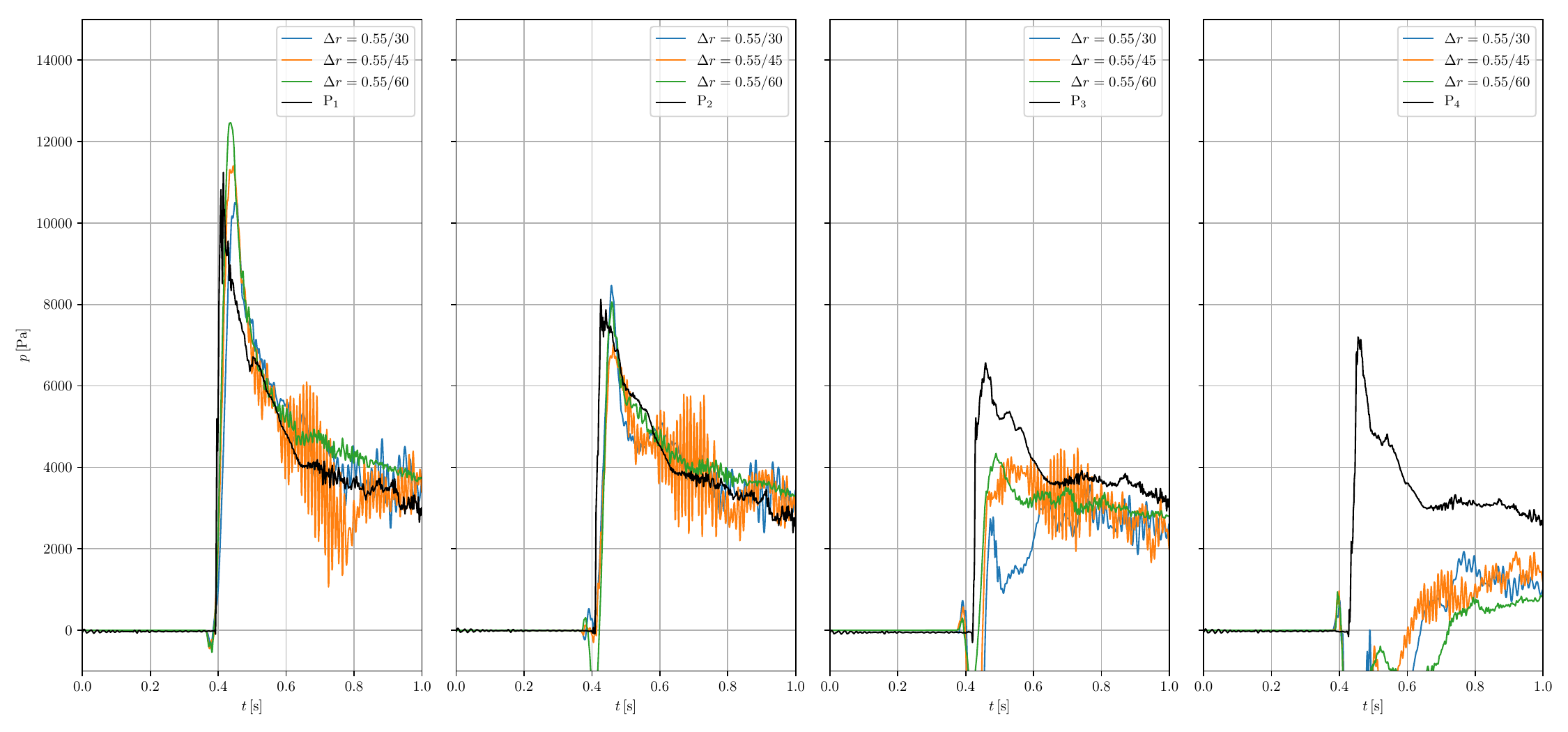}
\end{center}
\caption{Evolution of the reduced $L_2$ hydrostatic pressure error for the hydrostatic case. Solid lines: new formulation for three different resolutions ($10 000$, $50 000$ and $100 000$), dashed lines: traditional formulation.}
\label{fig:p_dr}
\end{figure}

\begin{figure}[t]
\begin{center}
\includegraphics[width=0.9\columnwidth]{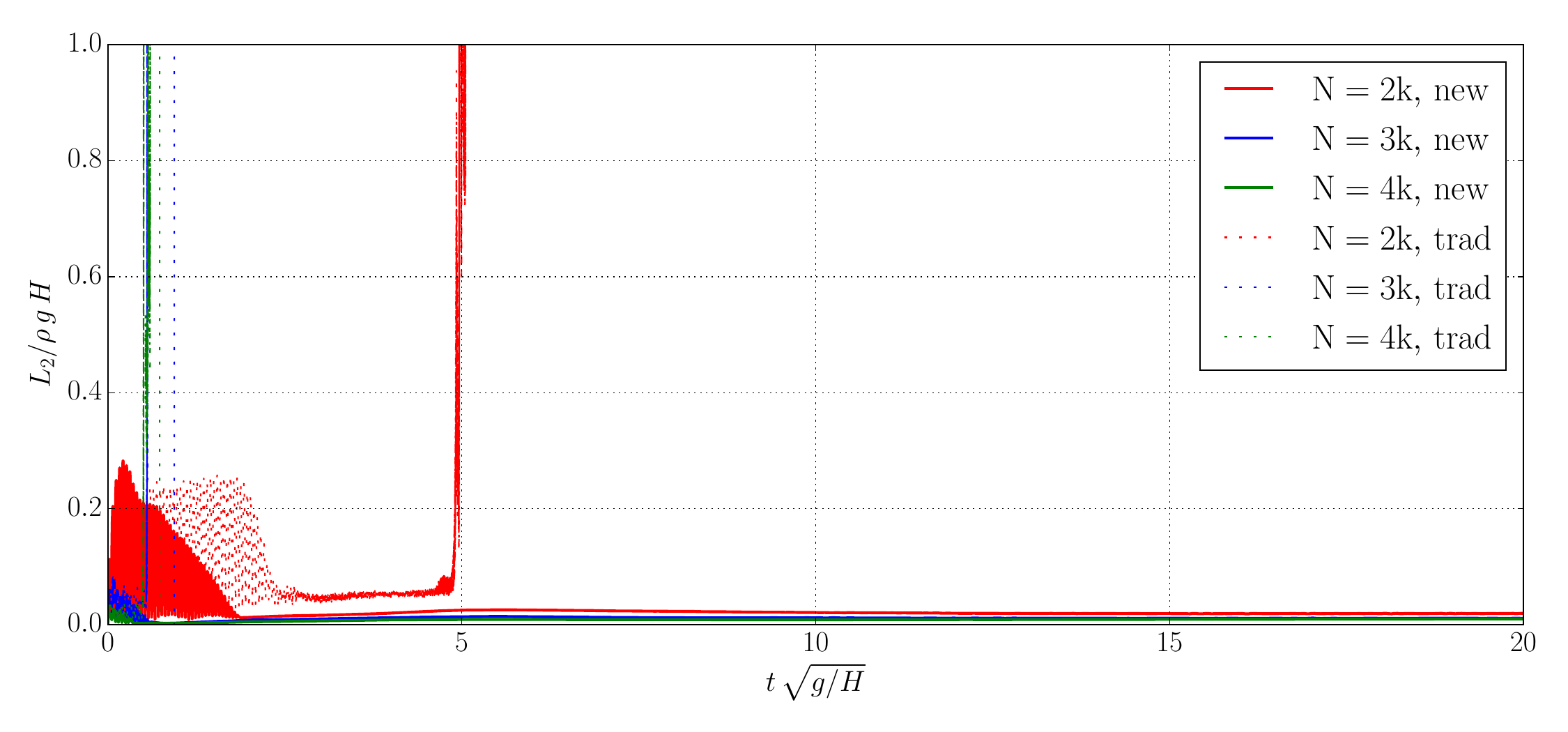}
\end{center}
\caption{Evolution of the non-dimensional $L_2$ hydrostatic pressure error for the hydrostatic case. Solid lines: new formulation for three different supports ($2$, $3$ and $4$),  dashed lines: traditional formulation for the same supports.}
\label{fig:p_dr_hfac}
\end{figure}

In order to  deeper analyze the differences between both formulations, Figure \ref{fig:kineticEnergy} presents the kinetic energy evolution for three different resolutions. As can be clearly appreciated, in the traditional formulation kinetic energy increases until the simulation fails. This time is shortened if the number of particles increases. Regarding the new formulation, the maximum value slightly increases as the number of particles increase, however its value is negligible and constant in time.

Figures \ref{fig:p_dr} and \ref{fig:p_dr_hfac} show the error in the computation of the hydrostatic pressure according to the  $L_2$ error of the pressure field (compared with the correct one), normalized by the maximum hydrostatic pressure $\rho g H$. Figure \ref{fig:p_dr} compared the traditional and the new formulations for three different resolutions, whilst Figure \ref{fig:p_dr_hfac} represents the same two formulations for the biggest resolution and three different supports.

In Figure \ref{fig:p_dr} results are seen to follow the same trends as the kinetic energy. Errors in the pressure field start to rise until the simulation fails. The failure occurs earlier for finer resolutions.
Conversely, the new formulation has a relatively constant error that diminishes as the resolution increases, being almost the same for the two finer resolutions tested.

The same behavior is found when testing different supports. Whilst the traditional formulation tends to increase the error when support increases, in the new formulation this change is almost not noticeable,  with a reasonable and constant error in time being obtained.

\subsection{3-D dam break}
\label{subsec:applications:dam-break}
In the previous application the capability of the new formulation to significantly improve the results, specially close to the contact line, has been checked.
However, in order to assess the performance of the new method in a more general context, the SPHERIC validation test number 2, consisting on a 3-D dam break flow, is here considered.
This validation test has already been considered in a similar context in Refs. \cite{Mayrhofer15b,violeau2014_spheric}.

\begin{figure*}[t]
	\centering
	\includegraphics[width=0.8\textwidth]{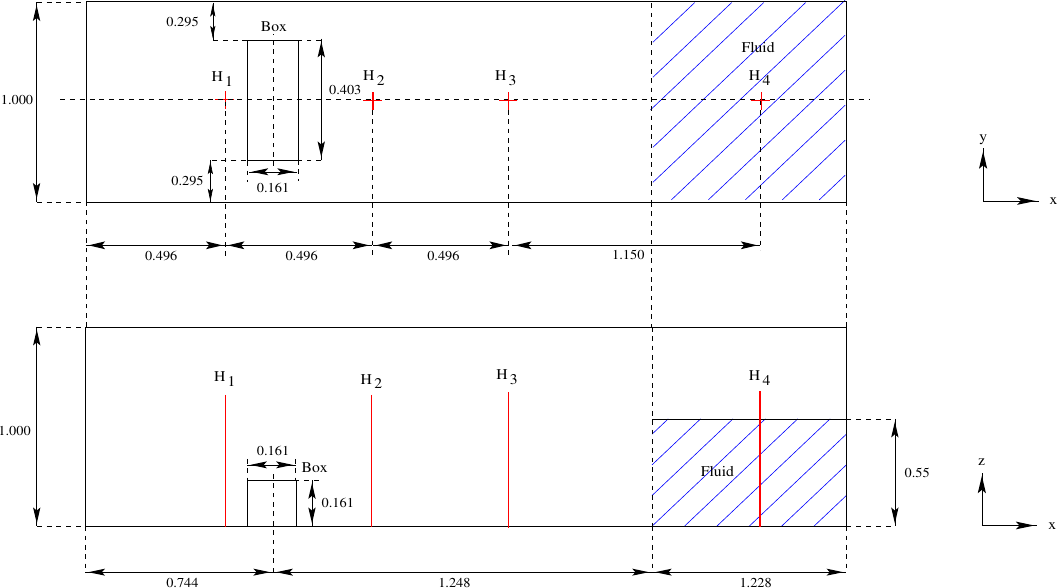}
	\caption{Schematic 3-D dam break flow initial condition \cite{kleefsman2005}}
	\label{fig:applications:dam-break:scheme}
\end{figure*}

The 3-D dam break initial condition is schematically depicted in Fig. \ref{fig:applications:dam-break:scheme}.
A reservoir of water with the shape of a rectangular box, with dimensions $1.228$ m $ \times 1 $ m $ \times 0.55$ m, is initially set in hydrostatic equilibrium at one side of the tank, which has dimensions $3.22 $ m$\times 1 $ m$ \times 1$ m.
A fixed box of dimensions $0.161 $ m$ \times 0.403$ m$ \times 0.161$m is also placed inside the tank.

\begin{figure}[t]
	\centering
	\includegraphics[width=0.45\textwidth]{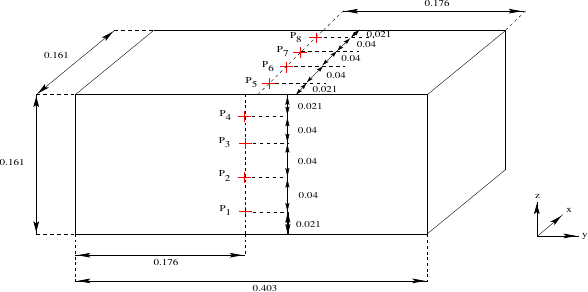}
	\caption{Schematic inner box description \cite{kleefsman2005}}
	\label{fig:applications:dam-break:scheme_box}
\end{figure}

Experiments on this test case have been described in Ref. \cite{kleefsman2005}.
In such experiments the wave height is quantitatively measured in the vertical probes labeled H$_1$-H$_4$.
An additional set of pressure sensors are distributed along the fixed inner box, as schematically depicted in Fig. \ref{fig:applications:dam-break:scheme_box}.

For the sake of simplicity, herein we focus in the first impact of the flow against the inner box, and more specifically in the pressure records measured in the sensors P$_1$-P$_4$. 

To carry out the simulations, a numerical sound speed of $c_0 = 40 $~m/s is considered, as well as the typical water density and viscosity values, $\rho_0 = 998$~kg/m$^3$ and $\mu = 10^{-6}$ m$^2$ / s.
The $\delta$-SPH formulation \cite{Antuono12} is applied for stability, following Ref. \cite{Fatehi11} (see also Ref. \cite{Cercos-Pita16a}, where several $\delta$-SPH term alternatives are discussed.)
This is in contrast with the simulations carried out in Refs. \cite{violeau2014_spheric} and \cite{Mayrhofer15b}, where a large artificial viscosity was considered in order to preserve the stability.
In these works, experimental records were not included, and SPH results were compared against a Volume-of-Fluid Finite Volume simulation.

No-slip boundary conditions have been imposed along the solid walls.
However, in this practical application a boundary force is added as well, to avoid walls penetration. It is triggered when a fluid particle moves closer than $0.1 \, \Delta r$ to the wall.
Even though such a boundary force has a detrimental effect in the pressure record noise, it is nevertheless required to avoid dramatically small time steps.

The air phase is neglected, for optimization purposes, assuming the inconsistencies close to the free surface discussed in Refs. \cite{Colagrossi09} and \cite{Colagrossi11} which have been also discussed previously in Section \ref{sec: Boundary Integrals}.
Besides the consistency issues close to the free surface, neglecting the air phase may induce some other errors in the pressure records, as already mentioned in Ref. \cite{Mayrhofer15b}.

In order to analyze the effect of the discretization on the results, several initial particle spacings have been considered, $\Delta r = 0.55 / 30$ m, $0.55 / 45 $ m, and $0.55 / 60$ m, with a constant kernel length ratio, $h / \Delta r = 4$.
Along the same line, several kernel lengths have been simulated, with a spacing of $\Delta r = 0.55 / 45$ m, namely $h / \Delta r = 2$, $3$ and $4$.

\begin{figure*}[t]
	\centering
	\includegraphics[width=0.9\textwidth]{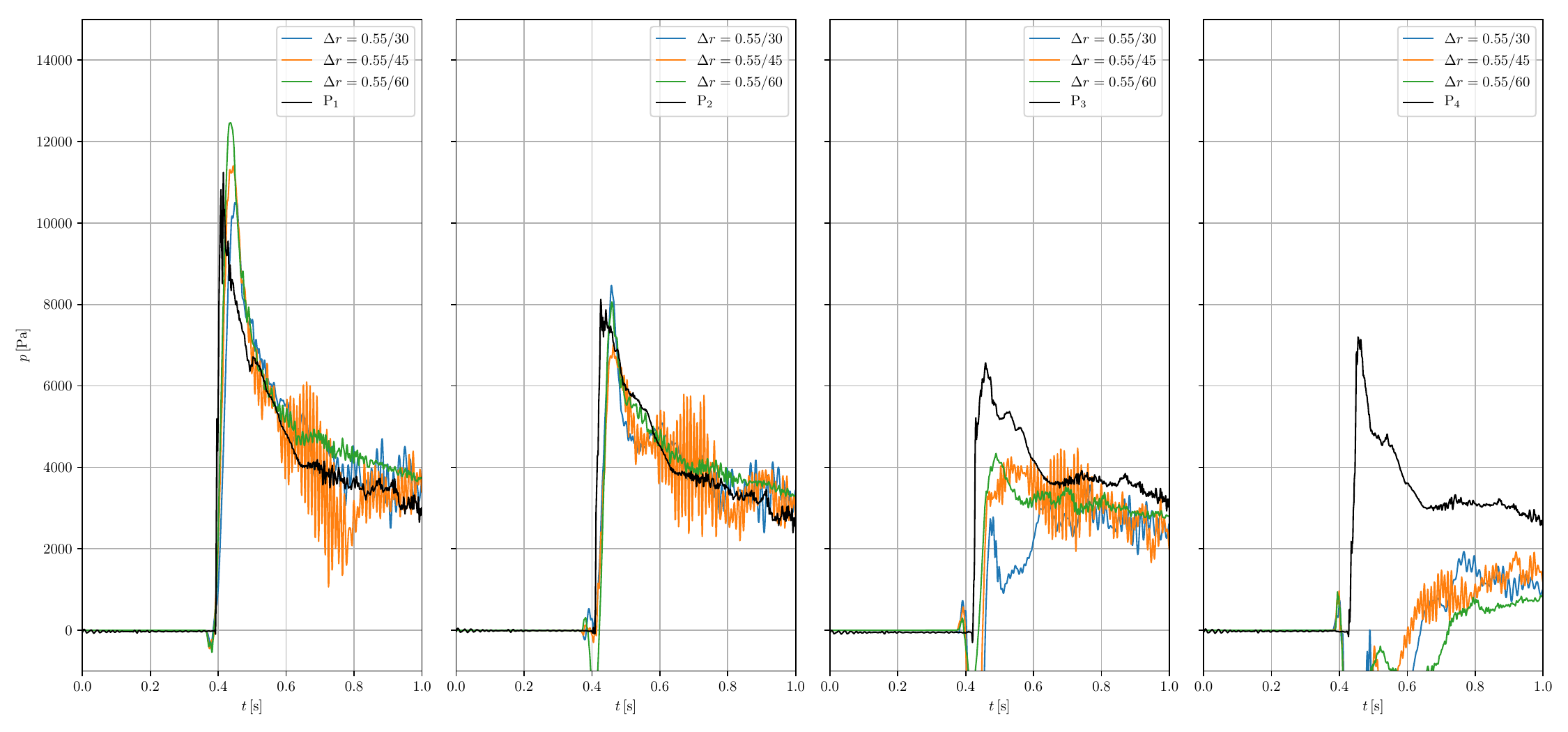}
	\caption{Pressure validation for different initial particle spacing values}
	\label{fig:applications:dam-break:p_dr}
\end{figure*}

In Fig. \ref{fig:applications:dam-break:p_dr}, the pressure records computed by AQUAgpusph \cite{Cercos-Pita15b} are compared with the experimental ones, for different initial particle spacing values.

Pressure records are obtained with a $1000$ FPS sampling rate. On top of that, high frequency noise has been filtered out by means of a Savitsky-Golay filter \cite{savitzky1964}.

After time $t = 0.5$ s, the pressure record associated to spacing $\Delta r = 0.55 / 45$ is significantly affected by a $\sim 100$ Hz frequency noise signal, whose source is not clear.

Nevertheless, a good agreement in the pressure measured at the pressure probes P$_1$ and P$_2$ is achieved for all the simulations, considering that SPH tends to overestimate the pressure peak at sensor P$_1$.
The same cannot be said for the pressure sensors P$_3$ and P$_4$, where the phenomenon is poorly captured in the simulation.

The overestimated pressure peaks, as well as the errors at P$_3$, have been attributed in the past to the absence of the air phase, see e.g. \cite{Mayrhofer15b}.
Unfortunately, the P$_4$ pressure record, which is consistently the one most affected by the neglect of the air phase, has been circumvented in the SPH literature \cite{Crespo2011,Lee2010,Mayrhofer15b,violeau2014_spheric}.
Anyway, given the accuracy of the records at probes P$_1$ and P$_2$, and the large errors at P$_3$ and P$_4$, it is plausible to conclude that the air phase plays an important role in the experimental pressure.

\begin{figure*}[t]
	\centering
	\includegraphics[width=0.9\textwidth]{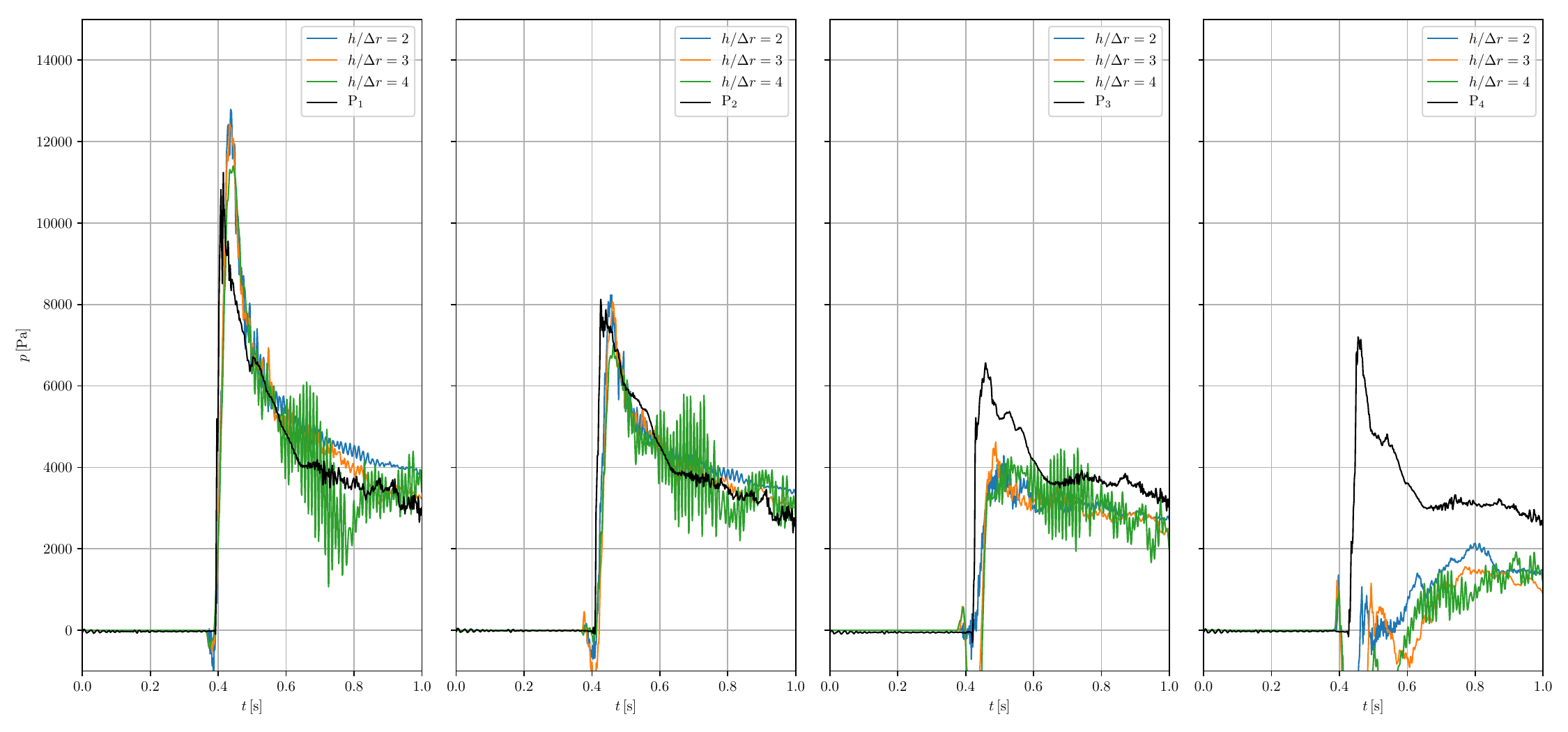}
	\caption{Pressure validation for different kernel length ratios}
	\label{fig:applications:dam-break:p_hfac}
\end{figure*}

In Fig. \ref{fig:applications:dam-break:p_hfac} similar pressure validations are shown for different kernel length ratios and constant initial particle space, $\Delta r = 0.55 / 45$.

The number of neighbors plays so far a secondary roll in the computed pressure at sensors  P$_1$-P$_2$.
However, the reduction of the kernel length, and therefore the smoothing radius, improves the pressure at P$_4$, specially close to the pressure peak, even if it always poorly captured in SPH.

Again, the simulation associated to the parameters $\Delta r = 0.55 / 45$ m and $h / \Delta r = 4$ is significantly affected by a $\sim 100$ Hz noise signal, which is not seen for the other lengths at all.

From the results discussed above, it is reasonable to suggest that simulations are correctly converging, both when the particles spacing is decreased and when the number of neighbors is increased.

As stated above, the computational performance can be a critical factor, as already already stressed in Ref. \cite{violeau2014_spheric}, where the computation of the analytical value of $\gamma$ took 40\% of the overall computation time.
In Table \ref{tab:applications:dam-break:computational_performance}, the time required to compute a single time step, in a NVIDIA GeForce GTX 750 Ti graphics device, is presented for all the simulations described above, for the traditional and the new $\gamma$ formulations.
Additionally, the performance for $\gamma = 1$, i.e. when the factor is not computed at all, has been added in order to be able to evaluate the overall impact.

\begin{table*}[!]
	\centering
	\label{tab:applications:dam-break:computational_performance}
	\caption{Time in seconds to compute a single time step, averaged along the first 100 time steps.}
	\begin{tabular}{c c  c c c}
		$\Delta r$ (m) &  $h / \Delta r$   & Former formulation & New formulation & $\gamma = 1$
		\\ \hline
		$ 0.55 / 30$& $ 4$ & $0.89$        & $0.84$     & $0.80$
		\\ 
		$ 0.55 / 45$& $ 4$ & $2.95$        & $2.82$     & $2.68$
		\\
		$ 0.55 / 60$& $ 4$ & $7.24$        & $6.93$     & $6.62$
		\\
		$ 0.55 / 45$& $ 2$ & $0.50$        & $0.48$     & $0.43$
		\\
		$ 0.55 / 45$& $ 3$ & $1.37$        & $1.28$     & $1.18$
	\end{tabular}
\end{table*}
The traditional $\gamma$ formulation takes $\sim 10\%$ of the overall computational time when large number of neighbors is considered, growing up to $\sim 16\%$ for the lowest number of neighbors, $h / \Delta r = 2$.
The new formulation requires almost half the time for the largest number of neighbors, i.e. $\sim 5\%$ of the overall computation time. Such number rises to $\sim 10\%$ for the lowest number of neighbors.
The general reduction of the computational cost, when the new formulation is applied, reflects the benefit of moving from a volume integral to a surface one.

\begin{figure}[t]
	\centering
	\includegraphics[width=0.45\textwidth]{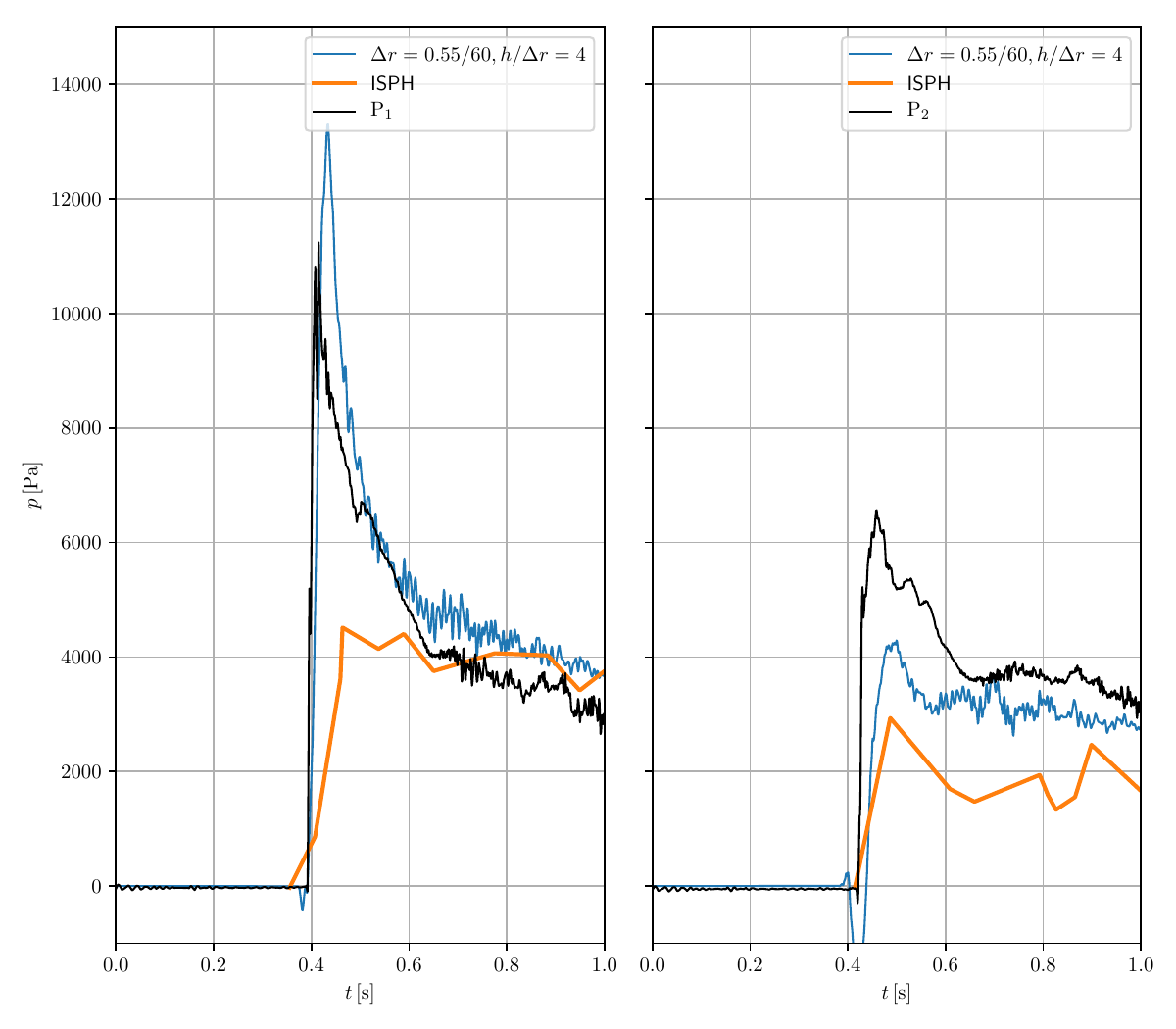}
	\caption{Pressure comparison with an ISPH solution}
	\label{fig:applications:dam-break:p_lee2010}
\end{figure}

For completeness, it is interesting to compare the results obtained with the new formulation, with the ones obtained in \cite{Lee2010}, where incompressible-SPH (ISPH) was applied.
In Fig. \ref{fig:applications:dam-break:p_lee2010} the pressure records at the sensors $\mathrm{P}_1$ and $\mathrm{P}_3$, for the ISPH simulation and the WCSPH simulation with the new formulation are depicted, and compared with the experimental results.
In this particular case, WCSPH has been able to capture significantly better the flow impact phenomena.

\section{Conclusions}
\label{5. Conclusions}

In the present work, a new kernel formulation is presented with the aim of reducing consistency issues for the boundary integrals formulation when free surfaces and boundaries are present in the interpolation. The kernel is defined in such a way that the whole volume integral is cast into a surface integral when boundaries are present, thus formulating the Shepard renormalization factor as a purely geometrical factor.

The expression which makes possible to write the Shepard factor as a function only of the geometry has a disadvantage: the kernel expression includes a divergent term which might cause problems when dealing with particles very close to the boundary. However, it is possible to split the expression into a numerical part, which is formed by a polynomial expression, and solve the divergent term analytically. The analytical term consists basically on a function of the subtended angle, which can be easily computed both in 2-D and 3-D

The new formulation has been tested against different planar boundaries, including corners, showing that the Shepard is computed correctly at the boundary. The new geometrical factor has been included into the tool AQUAgpusph and 
tested for two different applications. First, a 2-D hydrostatic stationary tank simulation is carried out in order to assess the performance of the new formulation and see if the boundary term computes properly. Errors in the consistency of the operators that lead to spurious velocities and wrong pressure values in the traditional formulation are now minimized. Different resolutions and compact supports have been tried and compared to the results obtained with the traditional formulation.
Second, a 3-D case is performed to show the applicability of the new formulation in a more general context, consisting on a 3-D dam break. The pressure value at four different points is presented and compared to other SPH formulations and the experimental results, showing a reasonable behavior for two of the probes and not as good for the other two, although this is in accordance with previous SPH simulations carrying out this test.
Also, computational performance is studied by comparing the time to compute a single time step that both formulations employ. Results show a general reduction in the computation time, reflecting the benefits of moving from a volume integral to a surface integral approach.

Overall, a potentially powerful formulation for computing the Shepard renormalization factor with application to the boundary integrals formulation is presented, showing that it can be easily implemented both in 2-D and 3-D and that is applicable to a wide range of engineering problems.

\section*{Acknowledgments}

The authors would like to thank Prof. Antonio Souto-Iglesias for his enthusiasm and the interesting discussions, and for his valuable help pushing this work forward.

This research has received funding from Universidad Polit\'ecnica de Madrid under a pre-doctoral scholarship.

\appendix

\section{Expressions for the divergent part of $F$}
\label{sec:appendix}

\begin{figure*}[t]
\centering
\includegraphics[width=0.45\textwidth]{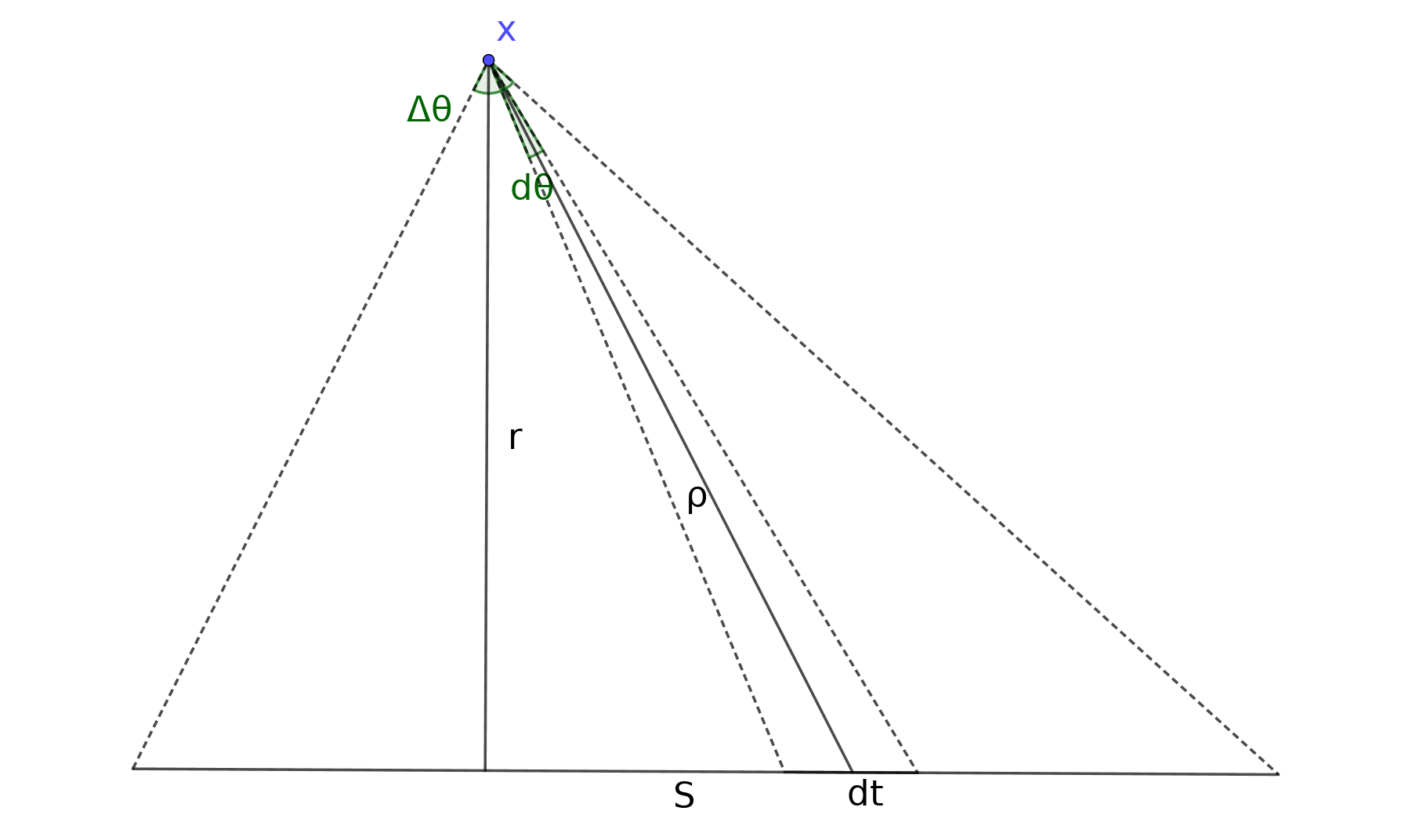}
\includegraphics[width=0.45\textwidth]{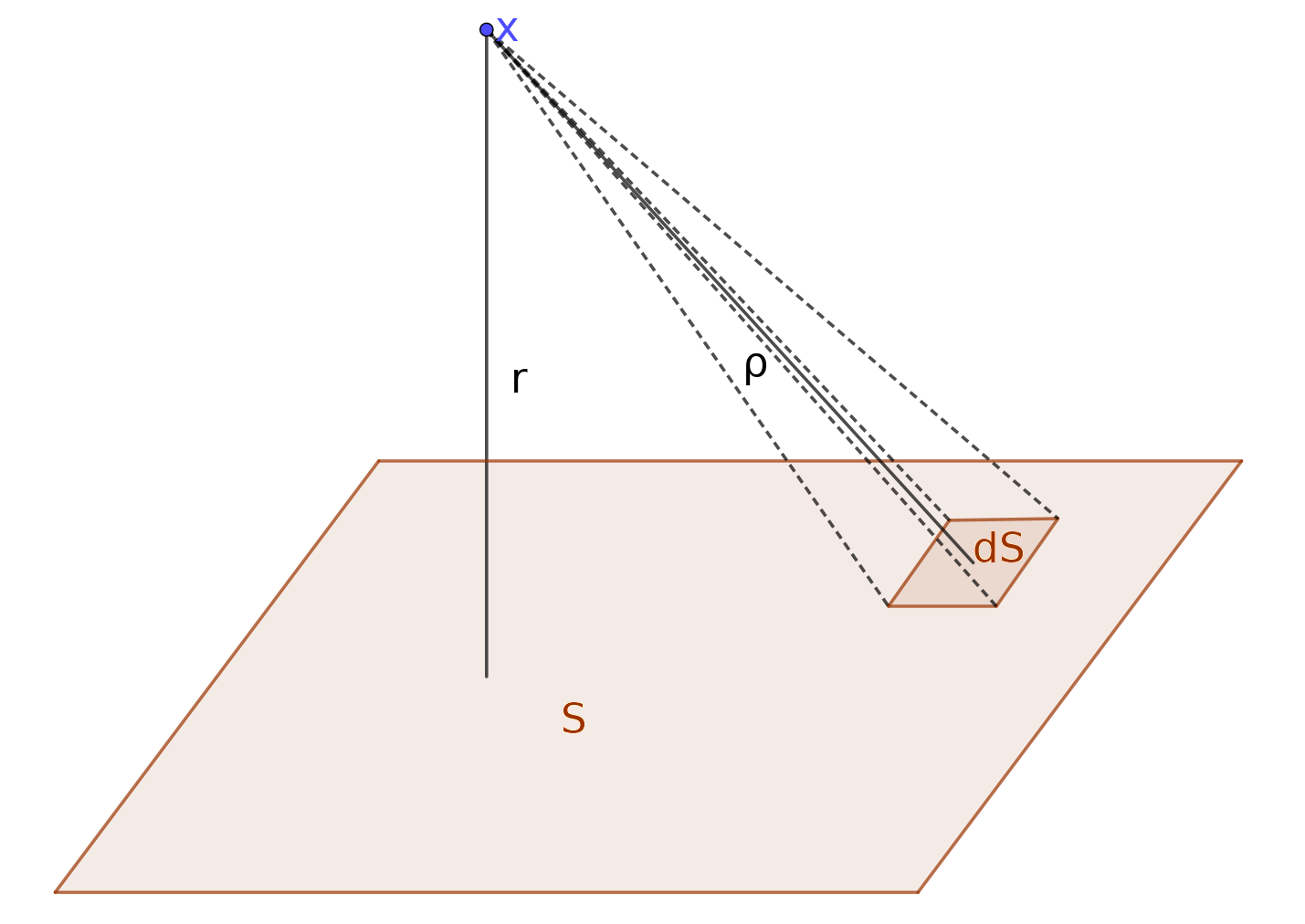}
\caption{Sketches of the integration over patches in 2-D (left)
  and 3-D (right)
    \label{fig:patches}}
\end{figure*}

It is worth mentioning that the expressions for the divergent part of $F$ ($F_\text{D}$), i.e. the last terms in Eqs. (\ref{eq:kernelSolution2D},~\ref{eq:kernelSolution3D}) are not limited to a particular kernel choice. Indeed, by its definition in Eq. \eqref{eq:kernelExpression}, $F$ is found by solving
\[
\left[ \rho'^d \, F (\rho') \right]_{\rho^d \, F (\rho)}^0 =
\int_\rho^{sh}
 \rho'^{d-1}  W_h (\rho') d\rho' ,
\]
where the integration limits guarantee that $F$ will have the same
support as $W_h$.

If we consider the $\rho\to 0$ limit, this yields
\[
- \rho^d \, F_\text{D} (\rho) =
\int_0^{sh}
\rho'^{d-1}  W_h (\rho') d\rho' .
\]
But, due to the kernel normalization, the right-hand side integral is $1/2\pi$ in 2-D, or  $1/4\pi$ in 3-D. From this fact, the same expressions for $F_\text{D}$ will arise in general. Notice that $\rho F_\text{D}(\rho)$ is basically the Coulomb field of a point particle, hence what is derived here closely follows the demonstration of Gauss' Law in electrostatics.

In Figure \ref{fig:patches} (left) the 2-D situation is sketched for a generic patch (index ``$j$'' will be dropped, for the sake of readability).  The distance $\bm{y} - \bm{x}$  can be decomposed in a constant normal distance to the patch ($r$), and a component tangent to the wall ($t$), which is the integration variable.

The integrals that appear in \eqref{eq:generalShepardSingular3} are of
the form
\[
\begin{split}
\bm{n} \cdot (\bm{y}_j - \bm{x} )
 \int_{S}    F (\bm{y} - \bm{x}) \, d\bm{y}  &=
 r
 \int_S    F (\rho ) \, dt = \\
 & =
 -\frac{r}{2\pi}
 \int_S \frac{1}{\rho^2}  \, dt .
\end{split}
\]
It is straightforward to demonstrate this equality,
\[
\frac{r}{\rho^2} dt = d\theta 
\]
(an expression which also appears in the calculations of the electrostatic field of a charged thin rod)
from which Eq. \eqref{eq:subtended_angle2D} results.

Figure \ref{fig:patches} (right) sketches the 3-D situation. In this case, the integrals have this form
\[
\begin{split}
 \bm{n} \cdot (\bm{y}_j - \bm{x} )
 \int_{S}    F (\bm{y} - \bm{x}) \, d\bm{y} &=
 r
 \int_S    F (\rho ) \, dS = \\
 & =
 -\frac{r}{4\pi}
 \int_S \frac{1}{\rho^3}  \, dt.
\end{split}
\]
In the same manner, we may demonstrate
\[
\frac{r}{\rho^3} dS = d\Omega ,
\]
this is, the differential of a solid angle. This way, Eq. \eqref{eq:subtended_angle3D} is obtained.

\section{Efficient evaluation of the angle subtended by a patch}
\label{sec:appendix2}
As discussed above, the alternative Shepard renormalization factor boils down to the computation of a convolution boundary term and an angle subtended by a boundary patch.
The former convolution boundary term, associated to the non-singular part of kernels in Eqs. (\ref{eq:kernelSolution2D},~\ref{eq:kernelSolution3D}), can be in fact computed applying the boundary integrals approach described in Refs. \cite{Ferrand13} and \cite{Cercos-Pita13}.
However, this is clearly not the case of the angle subtended by boundary patches.

As the boundary integrals methodology itself, the angle subtended by a boundary patch computation admits both a semi-analytical approach \cite{Ferrand13} and a purely numerical one \cite{Cercos-Pita13}, being the latter more efficient but less precise than the former.
Both approaches will be discussed here, in order to highlight their similarities and differences.
\subsection{Semi-analytical methodology}
\label{ssec:appendix2:semi-analytic}
\begin{figure*}[t]
	\centering
	\includegraphics[width=0.35\textwidth]{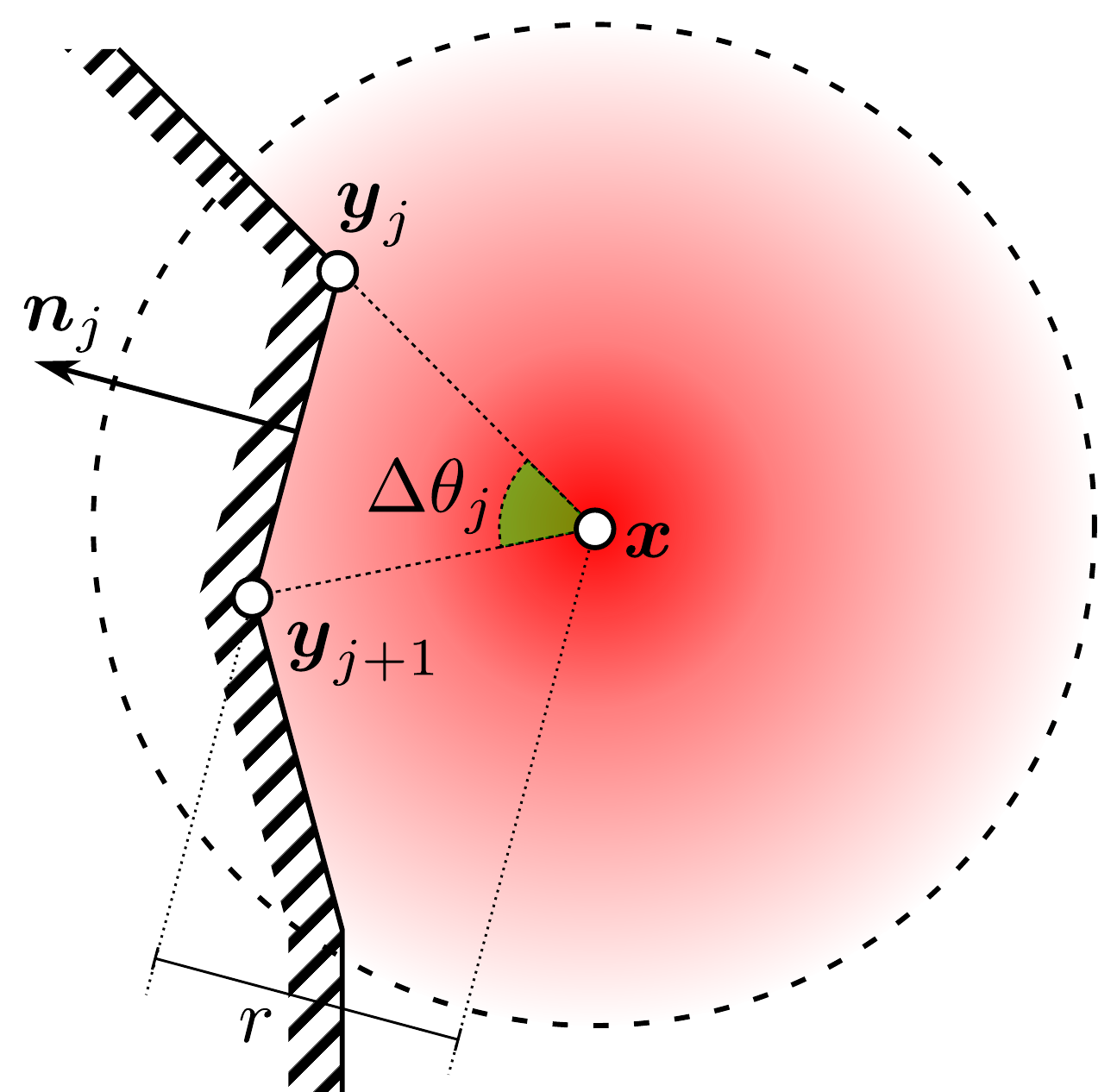}
	\includegraphics[width=0.35\textwidth]{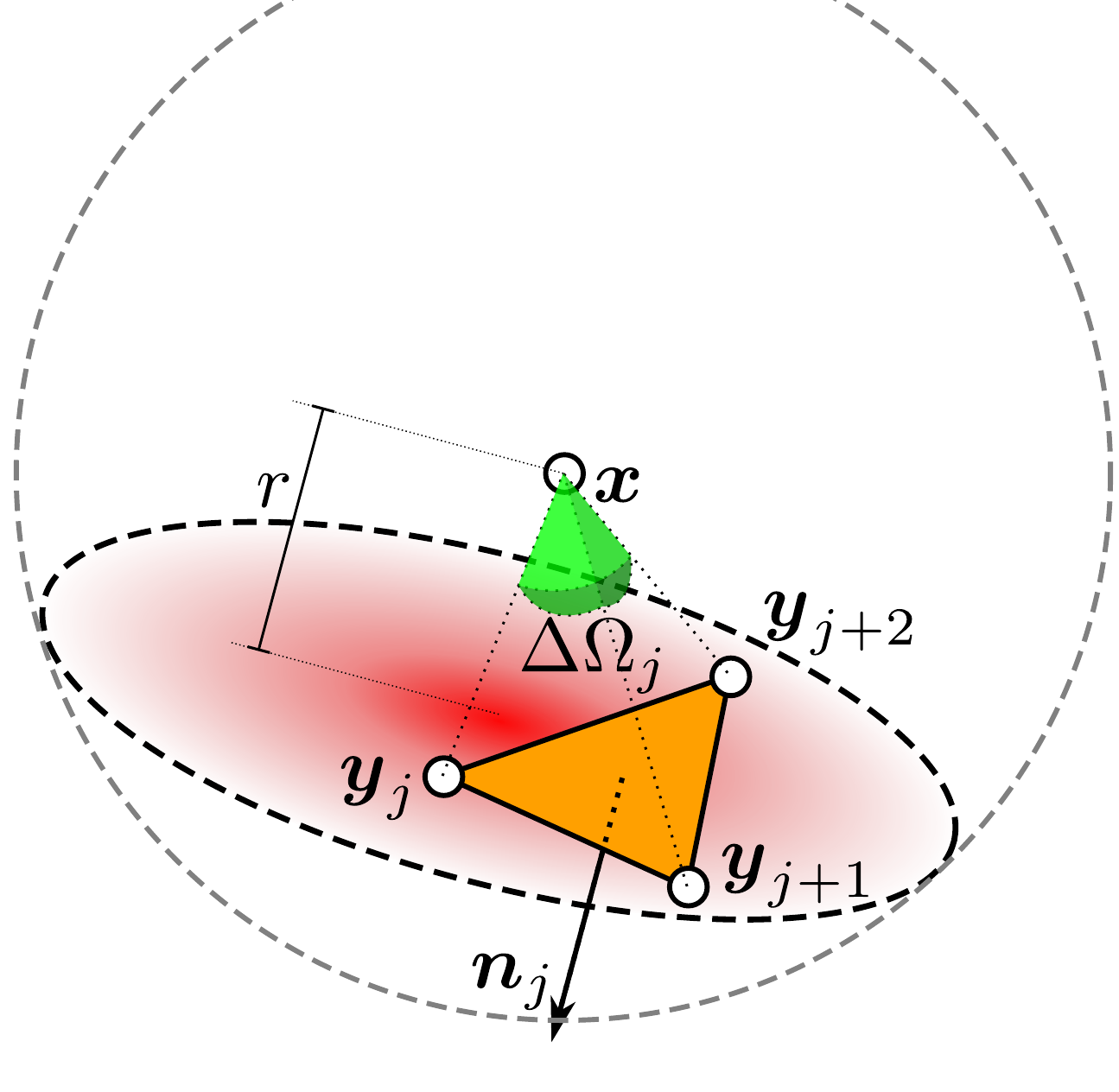}
	\caption{Schematic view of the subtended angle computation within the semi-analytical context \cite{Ferrand13}.}
	\label{fig:angle_semianalytic}
\end{figure*}

When the semi-analytical boundary integrals methodology is applied, as described in \cite{Ferrand13}, the boundary is discretized in planar patches, each defined by its vertices and normal vector, as it is schematically depicted in Fig. \ref{fig:angle_semianalytic}.

Indeed, in 2-D simulations the boundary is discretized in line segments, such that the tangent distance of a generic $j^\mathrm{th}$ vertex,
\[
r^t_j = \vert \left(\boldsymbol{y}_j - \boldsymbol{x} \right) - \boldsymbol{n}_j \left( \boldsymbol{n}_j \cdot \left(\boldsymbol{y}_j - \boldsymbol{x} \right) \right) \vert,
\]
can be used to compute the angle subtended by the generic $j^\mathrm{th}$ line segment:
\[
\Delta \theta_j = \tan^{-1} \left( \frac{r^t_{j+1}}{r} \right) + s_t \tan^{-1} \left( \frac{r^t_j}{r} \right),
\]
with
\[
s_t = \mathrm{sign}\left( \vert\boldsymbol{y}_j - \boldsymbol{x} \vert - \left( r^t_{j+1} - r^t_j \right) \right) .
\]

Even though 3-D boundaries can be discretized by a wide variety of planar patches, only triangulations will be discussed here. The expression to compute angles subtended by triangular plates, discussed in \cite{VanOosterom83}, can be applied:
\[
\left\lbrace
\begin{array}{l}
\boldsymbol{a} = \boldsymbol{y}_{j} - \boldsymbol{x}, \\
\boldsymbol{b} = \boldsymbol{y}_{j + 1} - \boldsymbol{x}, \\
\boldsymbol{c} = \boldsymbol{y}_{j + 2} - \boldsymbol{x}, \\
\Delta \Omega_j = \frac{\boldsymbol{a} \cdot \left(\boldsymbol{b} \times \boldsymbol{c}\right)}{
a b c +
a \left(\boldsymbol{b} \cdot \boldsymbol{c}\right) +
b \left(\boldsymbol{a} \cdot \boldsymbol{c}\right) +
c \left(\boldsymbol{a} \cdot \boldsymbol{b}\right)
}
\end{array}
\right.
\]
\subsection{Purely numerical methodology}

\label{ssec:appendix2:numeric}
In the purely numerical boundary integrals methodology \replaced[id=rev2]{(details of the implementation can be found in Refs. }{, discussed e.g. in Ref. \cite{Cercos-Pita13}}\cite{Cercos-Pita13,Cercos-Pita15b}), the same discretization applied to the volume particle is naturally extended to the boundary, i.e. the boundary is sampled by boundary elements associated to a boundary area portion.
Indeed, mesh connectivity between elements is not required any more, which permits a substantial improvement of computational performance.
On the other hand, truncation errors are introduced, meaning lower quality results.

\begin{figure*}[t]
	\centering
	\includegraphics[width=0.35\textwidth]{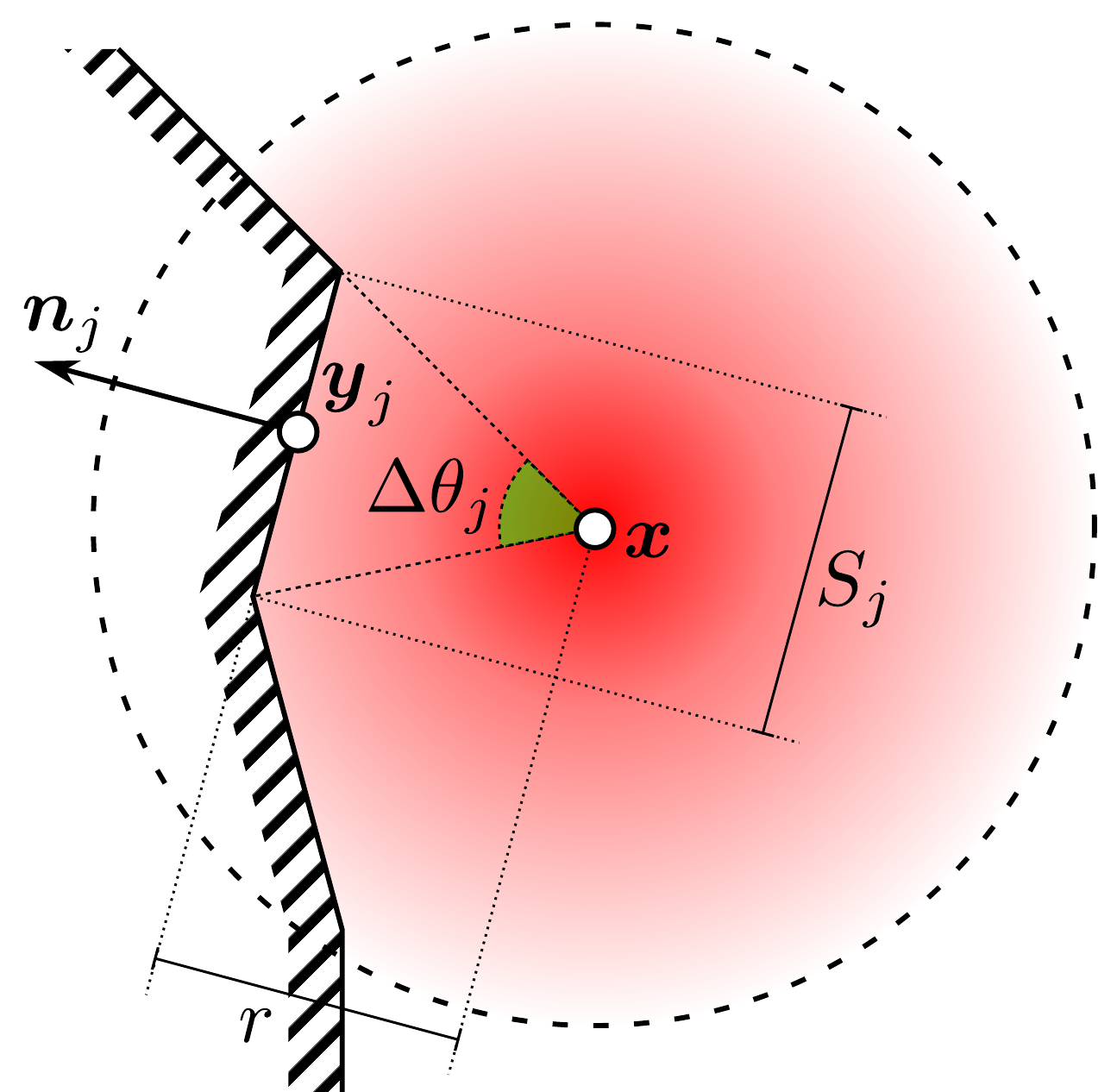}
	\includegraphics[width=0.35\textwidth]{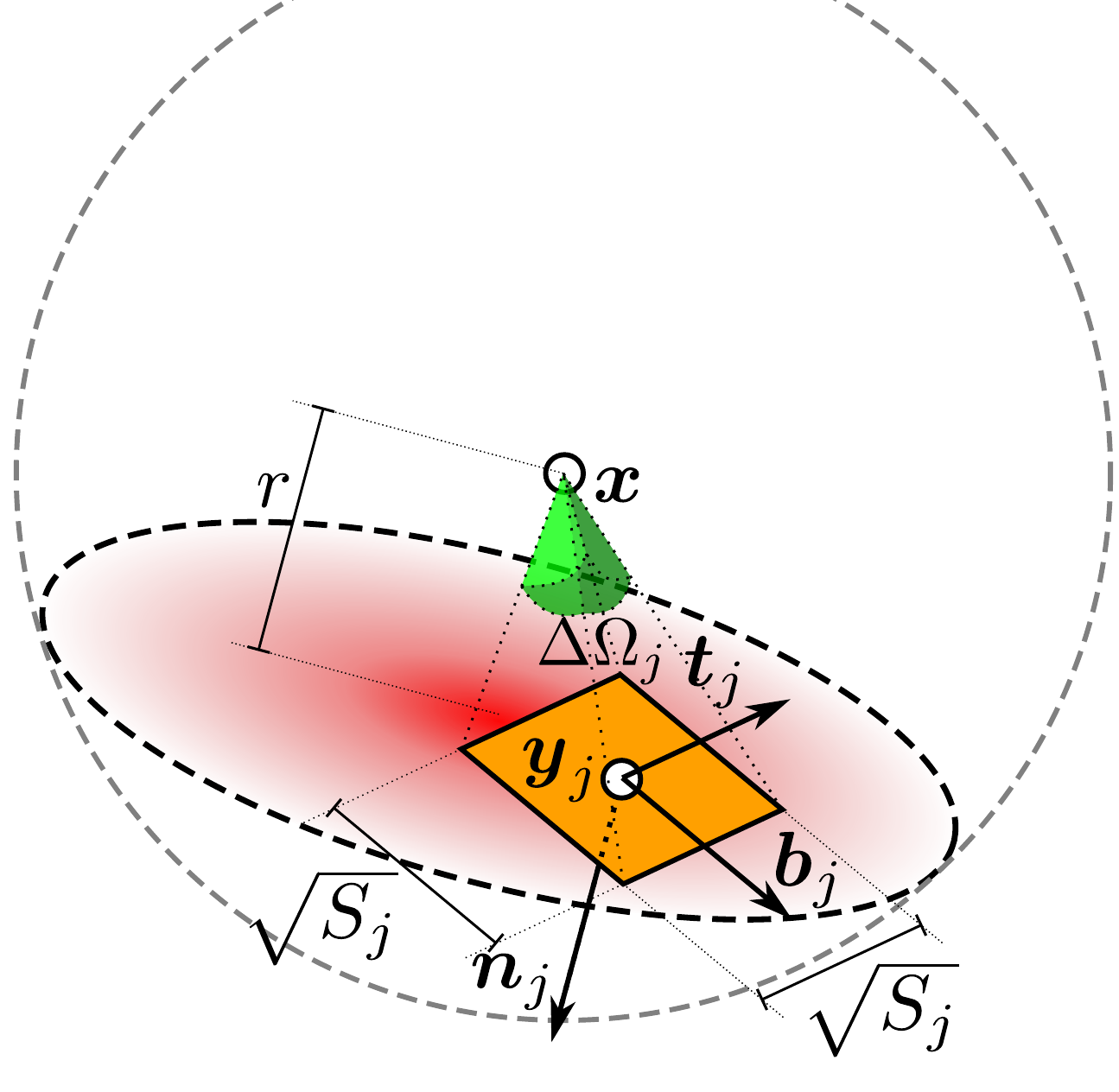}
	\caption{Schematic view of the subtended angle computation within the purely numerical context \cite{Cercos-Pita13}.}
	\label{fig:angle_numeric}
\end{figure*}

In fact, the 2-D implementation is rather similar to the semi-analytical one, discussed in \ref{ssec:appendix2:semi-analytic}.
The boundary is discretized in line segments that, in contrast to the semi-analytical approach, are defined by their center $\boldsymbol{y}_j$, area $S_j$ and normal $\boldsymbol{n}_j$, as schematically depicted in Fig. \ref{fig:angle_numeric} (left).
Analogously, we can compute the tangential distance to the generic  $j^\mathrm{th}$ boundary element center,
\[
r^t_j = \vert \left(\boldsymbol{y}_j - \boldsymbol{x} \right) - \boldsymbol{n}_j \left( \boldsymbol{n}_j \cdot \left(\boldsymbol{y}_j - \boldsymbol{x} \right) \right)\vert,
\]
with an expression for the angle subtended by the boundary element similar to the one already found for the semi-analytical methodology,
\[
\Delta \theta_j = \tan^{-1} \left( \frac{r^t_{j} + S_j / 2}{r} \right) - \tan^{-1} \left( \frac{r^t_j - S_j / 2}{r} \right).
\]

The 3-D case is much more complex. The boundary is then discretized in
square area elements, defined by their center $\boldsymbol{y}_j$, area
$S_j$, normal $\boldsymbol{n}_j$ as schematically depicted in
Fig. \ref{fig:angle_numeric} (right). \added[id=rev2]{A tangent vector
  $\boldsymbol{t}_j$ is considered, whose choice does not affect the
  results, as long as it is contained in the plane of the square.}
To compute the solid angle subtended by the boundary element we can make use of the well known expression of the solid angle subtended by a rectangular patch in which one corner coincides with the projection of the origin, $\boldsymbol{x}$, onto the plane:
\[
\Omega \left(a, b, r\right) = \cos^{-1} \left( \sqrt{ \frac{1 + (a / r)^2 + (b / r)^2}{(1 + (a / r)^2) (1 + (b / r)^2)} } \right),
\]
with $a$ and $b$ being the width and height of the rectangular patch respectively.

To this end, we can compute the tangential distances,
\[
\begin{array}{lcl}
r^t_j & = & \vert \boldsymbol{t}_j \cdot \left(\boldsymbol{y}_j - \boldsymbol{x} \right) \vert,
\\
r^b_j & = & \vert \boldsymbol{b}_j \cdot \left(\boldsymbol{y}_j - \boldsymbol{x} \right) \vert,
\end{array}
\]
where $\boldsymbol{b}_j$ is the binormal vector, $\boldsymbol{n}_j \times \boldsymbol{t}_j$.
The subtended solid angle is subsequently computed as follows:
\[ 
\begin{array}{lll}
\Delta \Omega_j & = & \Omega \left(r^t_j + \frac{\sqrt{S_j}}{2}, r^b_j + \frac{\sqrt{S_j}}{2}, r\right) \\
& - s_t & \Omega \left(r^t_j - \frac{\sqrt{S_j}}{2}, r^b_j + \frac{\sqrt{S_j}}{2}, r\right) \\
& - s_b & \Omega \left(r^t_j + \frac{\sqrt{S_j}}{2}, r^b_j - \frac{\sqrt{S_j}}{2}, r\right) \\
& + s_t \, s_b & \Omega \left(r^t_j - \frac{\sqrt{S_j}}{2}, r^b_j - \frac{\sqrt{S_j}}{2}, r\right),
\end{array}
\]
with
\[
\begin{array}{lcl}
s_t & = & \mathrm{sign}\left( r^t_j - \frac{\sqrt{S_j}}{2} \right),
\\
s_b & = & \mathrm{sign}\left( r^b_j - \frac{\sqrt{S_j}}{2} \right).
\end{array}
\]

\section*{\refname}

\bibliographystyle{CaF}
\bibliography{phdbib}

\end{document}